%% file: main.tex
\documentclass[submission,copyright,creativecommons]{eptcs}
\providecommand{\event}{GCM 2020} 
\usepackage{underscore}           

\usepackage[T1]{fontenc}
\usepackage{amsmath,latexsym,amssymb}
\usepackage{adjustbox} 

\usepackage{graphicx}

\PassOptionsToPackage{hyphens}{url}\usepackage{hyperref}

\usepackage{pgfplots}
\usepackage{subcaption}
\usepackage{amssymb}
\usepackage{algorithm}
\usepackage{algorithmic}
\usepackage{listings}
\definecolor{codegreen}{rgb}{0,0.6,0}
\definecolor{codegray}{rgb}{0.5,0.5,0.5}
\definecolor{codepurple}{rgb}{0.58,0,0.82}
\definecolor{backcolour}{rgb}{0.9,0.9,0.9}
\lstdefinestyle{mystyle}{
  language=C,
  backgroundcolor=\color{backcolour},  
  keywordstyle=\bfseries,
  numberstyle=\tiny\color{codegray},
  basicstyle=\ttfamily\footnotesize,
  breakatwhitespace=false,         
  breaklines=true,                 
  captionpos=b,                    
  keepspaces=true,                 
  numbers=left,                    
  numbersep=5pt,                  
  showspaces=false,                
  showstringspaces=false,
  showtabs=false,                  
  tabsize=2
}
\usepackage{tikz}
\usetikzlibrary{shapes.misc}
\tikzset{>=latex}
\tikzset{every edge/.style={draw,thick}}

\tikzset{gp2 node/.style={draw, rounded rectangle, minimum height=.55cm, minimum size=4mm, thick}}
\tikzset{gp2 red node/.style={draw, rounded rectangle, minimum height=.55cm, minimum size=4mm, thick, fill=gp2red}}
\tikzset{gp2 blue node/.style={draw, rounded rectangle, minimum height=.55cm, minimum size=4mm, thick, fill=gp2blue}}
\tikzset{gp2 green node/.style={draw, rounded rectangle, minimum height=.55cm, minimum size=4mm, thick, fill=gp2green}}
\tikzset{gp2 grey node/.style={draw, rounded rectangle, minimum height=.55cm, minimum size=4mm, thick, fill=gp2grey}}
\tikzset{gp2 any node/.style={draw, rounded rectangle, minimum height=.55cm, minimum size=4mm, thick, fill=gp2pink}}

\tikzset{gp2 root/.style={draw, rounded rectangle, minimum height=.55cm, minimum size=4mm, thick, double, double distance=0.3mm}}
\tikzset{gp2 red root/.style={draw, rounded rectangle, minimum height=.55cm, minimum size=4mm, thick, double, double distance=0.3mm, fill=gp2red}}
\tikzset{gp2 blue root/.style={draw, rounded rectangle, minimum height=.55cm, minimum size=4mm, thick, double, double distance=0.3mm, fill=gp2blue}}
\tikzset{gp2 green root/.style={draw, rounded rectangle, minimum height=.55cm, minimum size=4mm, thick, double, double distance=0.3mm, fill=gp2green}}
\tikzset{gp2 grey root/.style={draw, rounded rectangle, minimum height=.55cm, minimum size=4mm, thick, double, double distance=0.3mm, fill=gp2grey}}
\tikzset{gp2 any root/.style={draw, rounded rectangle, minimum height=.55cm, minimum size=4mm, thick, double, double distance=0.3mm, fill=gp2pink}}

\tikzset{gp2 edge/.style={->,thick}}
\tikzset{gp2 red edge/.style={->,thick,gp2red}}
\tikzset{gp2 blue edge/.style={->,thick,performanceBlue}}
\tikzset{gp2 green edge/.style={->,thick,gp2green}}
\tikzset{gp2 dashed edge/.style={->,thick,dashed}}
\tikzset{gp2 any edge/.style={->,thick,gp2pink}}

\tikzset{gp2 bi edge/.style={-,thick}}
\tikzset{gp2 red bi edge/.style={-,thick,gp2red}}
\tikzset{gp2 blue bi edge/.style={-,thick,performanceBlue}}
\tikzset{gp2 green bi edge/.style={-,thick,gp2green}}
\tikzset{gp2 dashed bi edge/.style={-,thick,dashed}}
\tikzset{gp2 any bi edge/.style={-,thick,gp2pink}}

\tikzset{none/.style={}}

\pgfdeclarelayer{nodelayer}
\pgfdeclarelayer{edgelayer}
\pgfsetlayers{nodelayer,edgelayer,main}
\pgfplotsset{compat=1.14}

\usepackage{xcolor}
\definecolor{gp2green}{RGB}{69, 191, 156}
\definecolor{gp2blue}{RGB}{153, 187, 255}
\definecolor{gp2red}{RGB}{236, 107, 116}
\definecolor{gp2pink}{RGB}{239, 161, 193}
\definecolor{gp2grey}{RGB}{196, 192, 200}
\definecolor{performanceBlue}{RGB}{0, 136, 255}
\definecolor{performanceYellow}{RGB}{252, 199, 17}
\definecolor{performancePink}{RGB}{255, 0, 255}

\newenvironment{allintypewriter}{\ttfamily}{\par}

\input{macros} 

\title{A Fast Graph Program for Computing\\Minimum Spanning Trees}
\author{Brian Courtehoute and Detlef Plump
\institute{Department of Computer Science, University of York, York, UK}
\email{\{bc956,detlef.plump\}@york.ac.uk}
}
\def\titlerunning{A Fast Graph Program for Computing Minimum Spanning Trees}
\def\authorrunning{B. Courtehoute \& D. Plump}
\begin{document}
\maketitle

\begin{abstract}
When using graph transformation rules to implement graph algorithms, a challenge is to match the efficiency of programs in conventional languages. To help overcome that challenge, the graph programming language GP\texorpdfstring{\,}{ }2 features \emph{rooted} rules which, under mild conditions, can match in constant time on bounded degree graphs. In this paper, we present an efficient GP\texorpdfstring{\,}{ }2 program for computing minimum spanning trees. We provide empirical performance results as evidence for the program's subquadratic complexity on bounded degree graphs. This is achieved using depth-first search as well as rooted graph transformation. The program is based on Bo\-ruv\-ka's algorithm for minimum spanning trees. Our performance results show that the program's time complexity is consistent with that of classical implementations of Bo\-ruv\-ka's algorithm, namely $\text{O}(m \log n)$, where $m$ is the number of edges and $n$ the number of nodes.
\end{abstract}

\section{Introduction}
\label{sec:intro}

GP\texorpdfstring{\,}{ }2 is an experimental rule-based graph programming language with simple semantics to facilitate formal reasoning. It has been shown that every computable function on graphs can be expressed as a GP\texorpdfstring{\,}{ }2 program \cite{Plump17a}.

A challenge in rule-based graph programming is reaching the time efficiency of conventional programs due to the cost of graph matching. In general, finding a match for a graph $L$ in a graph $G$ takes $\text{size}(G)^{\text{size}(L)}$ time, when in practise, we often want to do it in constant time.

Other programming languages based on graph transformation rules include
    AGG \cite{Runge-Ermel-Taentzer11a},
    GReAT \cite{Agrawal-Karsai-Neema-Shi-Vizhanyo06a},
    GROOVE \cite{Ghamarian-deMol-Rensink-Zambon-Zimakova12a},
    GrGen.Net \cite{Jakumeit-Buchwald-Kroll10a},
    Henshin \cite{Arendt-Biermann-Jurack-Krause-Taentzer10a} and
    PORGY \cite{Fernandez-Kirchner-Mackie-Pinaud14a},
    but we are not aware that any of them are able to match the time complexity of subquadratic graph algorithms.

GP\texorpdfstring{\,}{ }2 allows to speed up graph matching by using \emph{rooted} graph transformation rules, which was first introduced by Bak and Plump \cite{Bak-Plump12a}. This enables nodes in the host graph declared as \emph{roots} to be accessed in constant time, making matching locally around those in constant time possible for connected graphs of bounded degree.

In previous work, we developed GP\texorpdfstring{\,}{ }2 programs that match the time complexity of their conventional counterparts on connected graphs of bounded degree. The first such program produces a 2-colouring, and was shown to match the measured execution times of a tailor-made 2-colouring C program \cite{Bak-Plump16a}. More GP\texorpdfstring{\,}{ }2 programs that run in linear time on connected graphs of bounded degree include tree recognition, binary DAG recognition, and topological sorting \cite{Campbell-Courtehoute-Plump-2019}.

\begin{figure}[ht]
\centering
\includegraphics{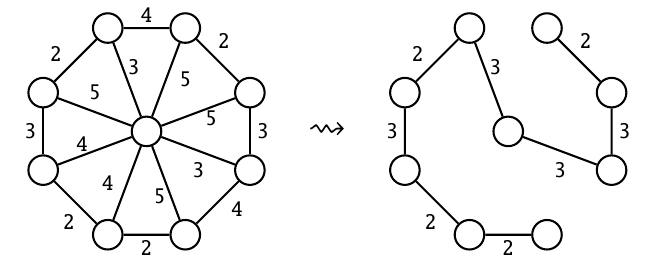}
\caption{A weighted graph and its minimum spanning tree}
\label{fig:mst-example}
\end{figure}

Here we continue this work by presenting an efficient GP\texorpdfstring{\,}{ }2 program computing a minimum spanning tree of a connected graph. Remember that a \emph{spanning tree} of an undirected connected graph $G$ with weighted edges is a subgraph that contains all nodes of $G$ and is a tree. A \emph{minimum spanning tree} (MST) of $G$ is a spanning tree such that the sum of all edge weights is minimum. For example, Figure \ref{fig:mst-example} shows a graph and its minimum spanning tree.

MSTs are useful for building networks between a set of nodes by minimising the cost. Such networks include communication, transport, piping, and computer networks. They also provide time efficient approximations to hard problems such as the travelling salesperson problem or the Steiner tree problem \cite{Skiena-2008}.

Classical algorithms for finding MSTs given by Prim, Kruskal, and Boruvka all run in time $\text{O}(m \log n)$, where $m$ is the number of edges and $n$ is the number of nodes \cite{Cunyet-Khali-2001}. However, to reach this time bound, the algorithms of Prim and Kruskal need data structures such as binary heaps or union-find data structures. In contrast, Boruvka's algorithm can be implemented efficiently without fancy data structures. Hence we choose to implement this algorithm in GP\texorpdfstring{\,}{ }2.

In Section \ref{sec:prog}, we give the GP\texorpdfstring{\,}{ }2 program \texttt{mst-boruvka}, which is based on depth-first search and rooted graph transformation. In Section \ref{sec:performance} we give execution time measurements as evidence that on bounded degree graphs, the program's complexity is consistent with the $\text{O}(m \log n)$ time bound of implementations of Boruvka's algorithm in conventional languages.

This paper is a revised and extended version of \cite{Courtehoute-Plump-2020}.

\section{The Graph Programming Language GP\texorpdfstring{\,}{ }2}
\label{sec:gp2}

This section briefly introduces GP\texorpdfstring{\,}{ }2, a graph transformation language, first defined in \cite{Plump12a}. Up-to-date versions of the syntax and semantics of GP\texorpdfstring{\,}{ }2 can be found in \cite{Bak15a}. The language is implemented by a compiler generating C code \cite{Bak-Plump16a,CampbellRomoPlump20}.
    
    \subsection{Graphs, Rules and Programs}
    \label{subsec:basics}
    
    GP\texorpdfstring{\,}{ }2 programs transform input graphs into output graphs, where graphs are directed and may contain parallel edges and loops. Both nodes and edges are labelled with lists consisting of integers and character strings. This includes the special case of items labelled with the empty list which may be considered as ``unlabelled''. 
    
    The principal programming construct in GP\texorpdfstring{\,}{ }2 consist of conditional graph transformation rules labelled with expressions. For example, the rule \texttt{min\_s} in Figure \ref{fig:mst-min} has three formal parameters of type \texttt{list}, two of type \ttt{int}, a left-hand graph and a right-hand graph which are specified graphically, and a textual condition starting with the keyword \texttt{where}.
    
    The small numbers attached to nodes are identifiers, all other text in the graphs consist of labels. Parameters are typed. In this paper we need the most general type \texttt{list} which represents lists with arbitrary values, and \ttt{int} which represents integers. 
    
    Besides carrying expressions, nodes and edges can be \emph{marked}\/ red, green or blue. In addition, nodes can be marked grey and edges can be dashed. For example, rule \texttt{root\_current} in Figure \ref{fig:mst-main} contains red and unmarked nodes and a red edge. Marks are convenient, among other things, to record visited items during a graph traversal and to encode auxiliary structures in graphs. The programs in the following sections use marks extensively.
    
    Rules operate on \emph{host graphs}\/ which are labelled with constant values (lists containing integers and character strings). Formally, the application of a rule to a host graph is defined as a two-stage process in which first the rule is instantiated by replacing all variables with values of the same type, and evaluating all expressions. This yields a standard rule (without expressions) in the so-called double-pushout approach with relabelling \cite{Habel-Plump02c}. In the second stage, the instantiated rule is applied to the host graph by constructing two suitable pushouts. We refer to \cite{Bak15a} for details and only give an equivalent operational description of rule application.
    
    Applying a rule $L \dder R$\/ to a host graph $G$\/ works roughly as follows: (1) Replace the variables in $L$ and $R$\/ with constant values and evaluate the expressions in $L$ and $R$, to obtain an instantiated rule $\hat{L} \dder \hat{R}$. (2) Choose a subgraph $S$\/ of $G$\/ isomorphic to $\hat{L}$ such that the dangling condition and the rule's application condition are satisfied (see below). (3) Replace $S$\/ with $\hat{R}$\/ as follows: numbered nodes stay in place (possibly relabelled), edges and unnumbered nodes of $\hat{L}$ are deleted, and edges and unnumbered nodes of $\hat{R}$ are inserted. 
    
    In this construction, the \emph{dangling condition}\/ requires that nodes in $S$\/ corresponding to unnumbered nodes in $\hat{L}$ (which should be deleted) must not be incident with edges outside $S$. The rule's application condition is evaluated after variables have been replaced with the corresponding values of $\hat{L}$, and node identifiers of $L$\/ with the corresponding identifiers of $S$. For example, the condition $\mtt{i<j}$ of rule \texttt{min\_s} in Figure \ref{fig:mst-min} requires that the integer label of the edge from node $g(\mtt{1})$ to node $g(\mtt{2})$ is smaller than the integer label of the edge from node $g(\mtt{1})$ to node $g(\mtt{3})$, where $g(\mtt{1})$, $g(\mtt{2})$, $g(\mtt{3})$ are the nodes in $S$\/ corresponding to \texttt{1}, \texttt{2}, \texttt{3}. 
    
    A program consists of declarations of conditional rules and procedures, and exactly one declaration of a main command sequence, which is a distinct procedure named \texttt{Main}. Procedures must be non-recursive, they can be seen as macros. We describe GP\texorpdfstring{\,}{ }2's main control constructs.
    
    The call of a rule set $\{r_1,\dots,r_n\}$ non-deterministically applies one of the rules whose left-hand graph matches a subgraph of the host graph such that the dangling condition and the rule's application condition are satisfied. The call \emph{fails}\/ if none of the rules is applicable to the host graph. 
    
    The command \texttt{if} $C$ \texttt{then} $P$ \texttt{else} $Q$ is executed on a host graph $G$ by first executing $C$ on a copy of $G$. If this results in a graph, $P$\/ is executed on the original graph $G$; otherwise, if $C$ fails, $Q$ is executed on $G$. The command \texttt{try} $C$ \ttt{then} $P$ \ttt{else} $Q$ has a similar effect, except that $P$\/ is executed on the result of $C$'s execution. If \ttt{then} $P$ or \ttt{else} $Q$ are omitted, no additional command is executed in the missing cases.
    
    The loop command $P!$ executes the body $P$\/ repeatedly until it fails. When this is the case, $P!$ terminates with the graph on which the body was entered for the last time. The \texttt{break} command inside a loop terminates that loop and transfers control to the command following the loop.
    
    In general, the execution of a program on a host graph may result in different graphs, fail, or diverge. The operational semantics of GP\texorpdfstring{\,}{ }2 defines a semantic function which maps each host graph to the set of all possible outcomes. See, for example, \cite{Plump17a}.
    
    \subsection{Rooted Programs}
    \label{subsec:rooted-programs}
    
    The bottleneck for efficiently implementing algorithms in a language based on graph transformation rules is the cost of graph matching. In general, to match the left-hand graph $L$\/ of a rule within a host graph $G$ requires time polynomial in the size of $L$ \cite{Bak-Plump12a,Bak-Plump16a}. As a consequence, linear-time graph algorithms in imperative languages may be slowed down to polynomial time when they are recast as rule-based programs. 
    
    To speed up matching, GP\texorpdfstring{\,}{ }2 supports \emph{rooted}\/ graph transformation where graphs in rules and host graphs are equipped with so-called root nodes. Roots in rules must match roots in the host graph so that matches are restricted to the neighbourhood of the host graph's roots. We draw root nodes using double circles. For example, in the rule \texttt{root\_current} of Figure \ref{fig:mst-main}, the nodes labelled \ttt{2} are roots and so is the node labelled \ttt{1} in the right-hand side.
    
    Rooted graph matching can be implemented to run in constant time under mild conditions, provided there are upper bounds on the maximal node degree and the number of roots in host graphs \cite{Bak-Plump12a}.

\section{Boruvka's Algorithm in GP\texorpdfstring{\,}{ }2}
\label{sec:prog}

In this section, we take a look at Boruvka's algorithm and its implementation in GP\texorpdfstring{\,}{ }2. We go through an example execution of the program \ttt{mst-boruvka} in Subsection \ref{subsec:example} in order to give an intuitive understanding of the program and how it relates to the algorithm. Subsections \ref{subsec:main}, \ref{subsec:preprocess}, \ref{subsec:findedge}, \ref{subsec:growforest}, and \ref{subsec:other} contain the program itself and its description.

Prim's, Kruskal's, and Bo\-ruv\-ka's algorithms for computing MSTs can all be implemented to run in $\text{O}(m \log n)$ time, where $m$ is the number of edges, and $n$ the number of nodes. However Prim's algorithm needs binary heaps to achieve it, and Kruskal's algorithm the union find data structure \cite{Cunyet-Khali-2001}. The advantage of Boruvka's algorithm is that it does not need fancy data structures to reach that time complexity bound \cite{Skiena-2008}. GP\texorpdfstring{\,}{ }2 has no predefined data structures except for the host graph that it transforms. Any additional data structures need to be encoded in the host graph itself, which can make a program tricky to read. Hence we choose to implement Boruvka's algorithm in GP\texorpdfstring{\,}{ }2.

Algorithm \ref{alg:boruvka} shows pseudocode for Boruvka's algorithm. Although it cannot translate directly into GP\texorpdfstring{\,}{ }2, it is a suitable starting point for the development of a GP\texorpdfstring{\,}{ }2 program.

\begin{algorithm}
\caption{Boruvka's MST algorithm on an input graph $G$}
\label{alg:boruvka}
\begin{algorithmic}[1]
\STATE \texttt{Preprocess}: initialise the spanning forest $F$ to be the nodes of $G$
\WHILE{$F$ consists of more than one tree}
    \FOR{each tree $T$ in $F$}
        \STATE \texttt{FindEdge}: select a minimum weight edge between $T$ and $G - T$, prioritising already selected edges if they are minimum
    \ENDFOR
    \STATE \texttt{GrowForest}: add the selected edges to $F$
\ENDWHILE
\end{algorithmic}
\end{algorithm}

The idea of Boruvka's algorithm is to initialise a forest as the nodes of the input graph without any edges, and to grow that forest by adding minimum-weight edges from between its connected components until it becomes a minimum spanning tree of the input graph.

\begin{figure}[ht]
\centering
\includegraphics{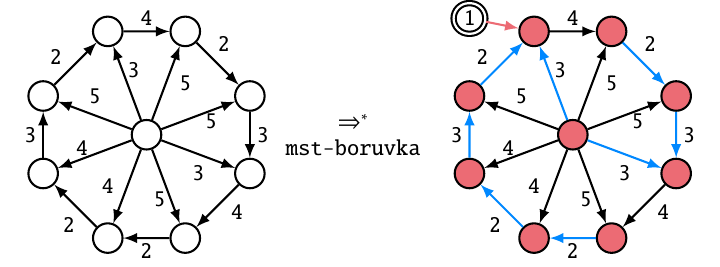}
\caption{Example input and output of \texttt{mst-boruvka}}
\label{fig:ex-wheel}
\end{figure}

As illustrated in Figure \ref{fig:ex-wheel}, the input of \ttt{mst-boruvka} is a connected graph with unmarked nodes and edges. Nodes are unlabelled, and edges have integer labels. In the output, the subgraph induced by the blue edges are a minimum spanning tree of the input. The additional root with label $1$ is an auxiliary construct used in the execution of the program (which could be removed in constant time).

\subsection{Example Execution}
\label{subsec:example}

\begin{figure}[ht]
\centering
\includegraphics{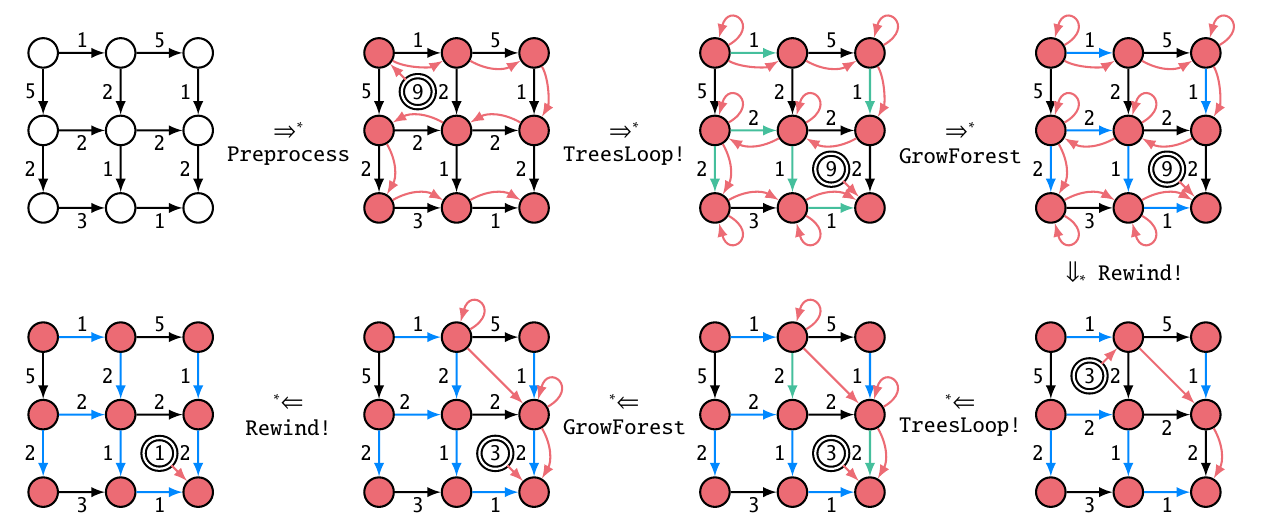}
\caption{Example execution of \texttt{mst-boruvka}}
\label{fig:ex-grid}
\end{figure}

Throughout the execution of \ttt{mst-boruvka}, the graph induced by the blue edges is a subgraph of the minimum spanning tree highlighted in the output. We shall call this forest $F$, and its connected components its \emph{trees}. Let us explore how \ttt{mst-boruvka} executes using the example in Figure \ref{fig:ex-grid}, and compare it to the pseudocode in Algorithm \ref{alg:boruvka}. The \ttt{Main} procedure of \ttt{mst-boruvka} is depicted in Figure \ref{fig:mst-main}.


The procedure \ttt{Preprocess} initialises the forest $F$ to be just the nodes of the input(see line 1 of the pseudocode). It also sets up a linked list of red edges and red nodes that helps the program loop over the trees of $F$ efficiently. Each tree of $F$ is represented by exactly one of its nodes being an entry in the linked list. Additionally, there is a pointer in the form of an unmarked root node with an outgoing red edge towards the ``current'' node in the linked list. The pointer also stores the number of trees the forest has in order to efficiently check whether only one tree is left, terminating the main loop (see line 2 of the pseudocode).

The loop \ttt{TreesLoop!} moves the pointer through the nodes of the linked list, effectively looping over the trees of $F$ (see line 3 of the pseudocode). On each tree $T$, the procedure \ttt{FindEdge} is called, which selects a minimum weight edge between $T$ and its complement in the host graph by marking it green (see line 4 of the pseudocode). If there is already an adjacent green edge with minimum weight, no new edge is selected since that could introduce a cycle into $F$. To ensure that only one node of each tree is part of the list, the current tree gets marked for deletion from the list using a red loop under certain conditions. Subsection \ref{subsec:other} elaborates on this.

The procedure \ttt{GrowForest} adds the selected edges to $F$ by green edges into blue ones (see line 6 of the pseudocode).

The loop \ttt{Rewind!} serves to maintain the linked list. It moves the pointer back to the beginning of the list. On the way, it removes nodes that have been marked for deletion with a red loop. It also decrements the pointer's label each time it encounters such a node since that node's tree has been merged with another tree.

\subsection{The GP\texorpdfstring{\,}{ }2 Program \ttt{mst-boruvka}}
\label{subsec:main}

\begin{figure}[ht]
		
\fbox{
\begin{minipage}{.97\textwidth}

\begin{allintypewriter}
Main = Preprocess; Loop!

\smallskip
Loop = if one\_tree then break else Body

\smallskip
Body = TreesLoop!; GrowForest; Rewind!

\smallskip
TreesLoop = root\_current; TraverseTree; MarkForDeletion; CleanUp;

\phantom{............}try next\_tree else break

\smallskip
TraverseTree = ColourBlue; FindEdge

\smallskip
CleanUp = ColourRed; unroot\_red!

\medskip

\setlength{\tabcolsep}{16pt}
\begin{tabular}{  p{6cm}  p{6cm}  }
				
	one\_tree () & root\_current (x,y:list)\\ 
	
	\adjustbox{valign=t}{\includegraphics{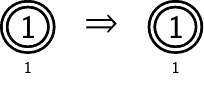}}
	&
	
	\adjustbox{valign=t}{\includegraphics{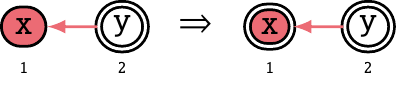}}
	\\
	
	\vspace{-1mm} next\_tree (x,y,z:list) & \vspace{-1mm} unroot\_red (x:list) \\  
	
	\adjustbox{valign=t}{\includegraphics{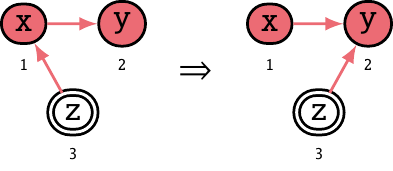}}
	& 
	
	\adjustbox{valign=t}{\includegraphics{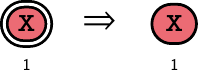}}

\normalsize
\end{tabular}
\end{allintypewriter}

\end{minipage}
}

\caption{The GP\texorpdfstring{\,}{ }2 program \texttt{mst-boruvka}}
\label{fig:mst-main}
\end{figure}

The program \ttt{mst-boruvka} is depicted in Figure \ref{fig:mst-main}. Most of it has been explained by the example execution in Subsection \ref{subsec:example}. Let us now examine the loop \ttt{TreesLoop!}.

The purpose of the loop \ttt{TreesLoop!} is to find a minimum weight edge from each tree to its complement and mark it green. It initialises by rooting the node the pointer points to. Then that node's tree is marked blue with the procedure \ttt{ColourBlue} so it can easily be distinguished from the rest of the graph. \ttt{FindEdge} then finds the minimum edge from the tree to its complement. The procedure \ttt{MarkForDeletion} marks the tree for deletion if it will be merged with another one. The procedure \ttt{ColourRed} makes the nodes of the tree be red again. The command \ttt{unroot\_red!} unroots any red roots. The rule \ttt{next\_tree} then moves the pointer to the next entry in the linked list.

\subsection{The Procedure \ttt{Preprocess}}
\label{subsec:preprocess}

\begin{figure}[ht]
		
\fbox{\begin{minipage}{.98\textwidth}

\begin{allintypewriter}
Preprocess = pre\_init; PreLoop!; unroot\_red

\smallskip
PreLoop = PreForward!; try pre\_back else break

\smallskip
PreForward = \string{pre\_forward1, pre\_forward2\string}

\medskip
\setlength{\tabcolsep}{8pt}
\begin{tabular}{  p{7cm}  p{7cm}  }
			
	pre\_init (x:list) & pre\_back (a,x,y:list) \\ 
	
	\adjustbox{valign=t}{\includegraphics{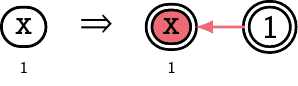}}
	&
	
	\adjustbox{valign=t}{\includegraphics{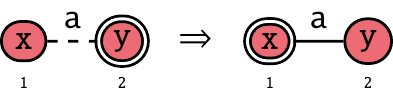}}
	\\

	\vspace{-1mm} pre\_forward1 (i:int; a,x,y:list) & \vspace{-1mm} pre\_forward2 (i:int; a,x,y,z:list) \\  
	
	\adjustbox{valign=t}{\includegraphics{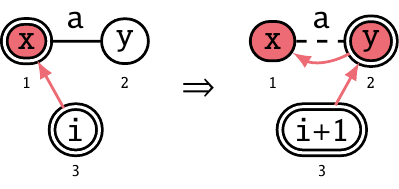}}
	& 
	
	\adjustbox{valign=t}{\includegraphics{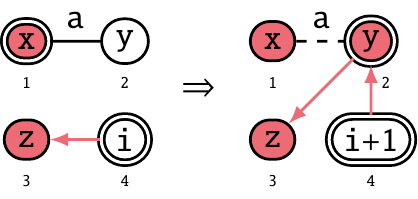}}

\normalsize			
\end{tabular}
\end{allintypewriter}

\end{minipage}
}

\caption{The procedure \texttt{Preprocess}}
\label{fig:mst-preprocess}
\end{figure}

The procedure \ttt{Preprocess} depicted in Figure \ref{fig:mst-preprocess} uses depth-first search (DFS) to construct the linked list and the pointer. An example of its input and output can be seen in Figure \ref{fig:ex-grid}.

The rule \ttt{pre\_init} initialises some node of the input to be the starting point of the DFS, and constructs the pointer. Since initially each node is its own tree, the pointer's label will count the number of nodes encountered during the DFS. Red nodes are considered to be discovered by the DFS, and unmarked nodes undiscovered.

The rules \ttt{pre\_forward1} and \ttt{pre\_forward2} are called non-deterministically. They both move the red root to an adjacent unmarked node. The rules contain \emph{bidirectional edges} (without arrowheads) that can be matched in either orientation. The rule is a shorthand for a non-deterministic call of copies of the same rule whose bidirectional edges have been replaced with directed edges in all possible combinations of orientation. The dashed edge serves as a way to keep track of the path the DFS has taken, which is backtracked by the rule \ttt{pre\_back}. The backtracking enables the ``forward'' rules to find new undiscovered nodes again.

The rules \ttt{pre\_forward1} and \ttt{pre\_forward2} also increment the counter and construct the linked list of red edges. The reason we need both rules is to cover both cases of whether the newest entry of the list is also the current red root or not.
 
\subsection{The Procedure \ttt{FindEdge}}
\label{subsec:findedge}

The procedure \ttt{FindEdge}\footnote{Compared to \cite{Courtehoute-Plump-2020}, we added several ``\texttt{min}'' rules to cover all cases.}, depicted in Figures \ref{fig:mst-findedge} and \ref{fig:mst-min}, serves to find a minimum-weight edge between the current tree (blue nodes) to the rest of the graph (red nodes) using DFS, and to mark it green. If among said minimum edges is an already selected (green) one, it will stay selected, and no additional edge is selected for the current tree. If this were not the case, the selected edges would form a cycle on a 3-cycle whose edges have equal weight for instance, causing the output MST not to be a tree.

\begin{figure}[ht]
\centering

\includegraphics{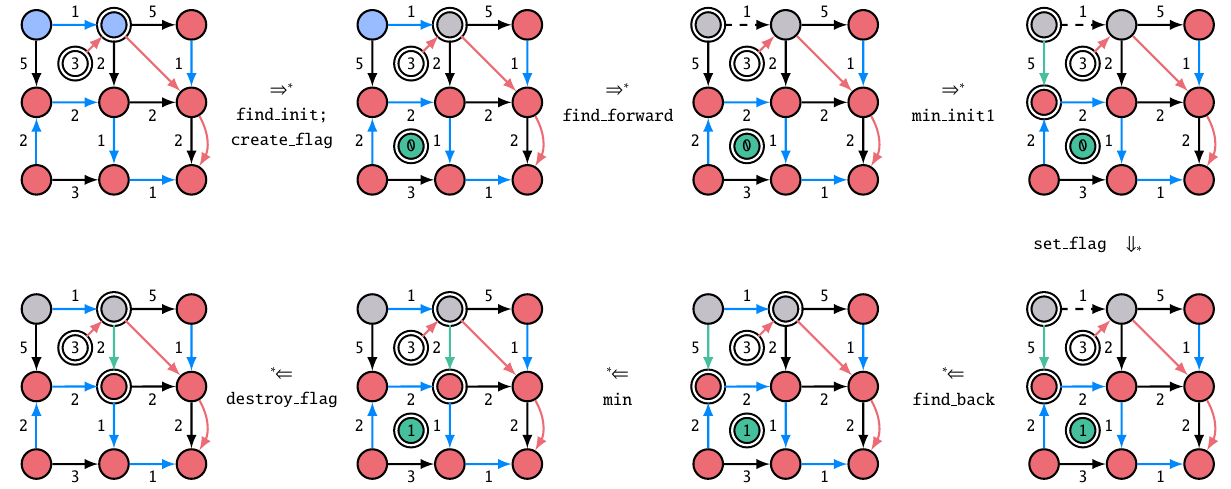}

\caption{Example execution of \texttt{FindEdge}}
\label{fig:ex-findedge}
\end{figure}

Let us examine the example execution of \ttt{FindEdge} in Figure \ref{fig:ex-findedge}. It is part of the transition from the fifth to the sixth graph labelled \ttt{TreesLoop!} in the example execution of \ttt{mst-boruvka} in Figure \ref{fig:ex-grid}. We start with a graph where the current tree has blue nodes to distinguish it from the rest of the graph. This was done using the procedure \ttt{ColourBlue}, which is always called before \ttt{FindEdge} as defined in the procedure \ttt{TraverseTree} in Figure \ref{fig:mst-main}. The nodes of the tree are turned grey, but are still distinguishable from the the rest of the graph, which has red nodes.

The procedure \ttt{FindEdge} starts by turning the blue root grey, and creating a green root which serves as a flag indicating whether the minimum edge has been initialised yet. The flag is \ttt{1} if initialisation has already happened, and \ttt{0} otherwise.

We enter the loop \ttt{FindLoop!} and apply \ttt{find\_forward} to move the root along in the current tree in a depth-first fashion. The flag is not yet set to \ttt{1}, so we call \ttt{MinSetup} to initialise the minimum edge using \ttt{min\_init1}. The rule \ttt{min\_init2} exists in case the grey root's only incident edge has already been selected (marked green) when the procedure \ttt{FindEdge} was applied to a different tree. An edge selected from both this tree and another tree is represented with a label that is a list consisting of the edge weight followed by a \ttt{0}. The currently selected minimum edge of the current tree is represented by a green edge incident to a grey as well as a red root.

We then enter the procedure \ttt{Success} which minimises the weight among the unmarked edges incident to the current grey root using the procedure \ttt{MinWithS} (which only calls rules that minimise edges incident to the current grey root), and then applies \ttt{set\_flag} to indicate that the initialisation of the minimum edge is complete.

Next, the rule \ttt{find\_back} moves the grey root back through the tree in depth-first fashion. We then enter the next iteration of \ttt{FindLoop!}. The rule \ttt{find\_forward} cannot be applied, so we continue with the loop \ttt{Minimise!} since the flag has already been set.

The purpose of the loop \ttt{Minimise!} is to find an edge incident to the grey root with a smaller weight than the currently selected edge. There are $14$ different cases we have to distinguish with the rules that update the minimum edge. 
They can be seen as combinations of the presence or absence of four flags \ttt{s}, \ttt{t}, \ttt{n}, and \ttt{p}, present in the rule names. The flag \ttt{s} is present if the new and previous minimum edge share their ``source'', i.e. the incident grey node in the current tree. The flag \ttt{t} is present if the new and previous minimum edge share their ``target'', i.e. the incident red node outside of the current tree. The flag \ttt{n} denotes that the new minimum edge is also a selected minimum edge of a different tree from a previous call of \ttt{FindEdge}. The presence of flag \ttt{p} indicates that the previous minimum edge has already been selected for a different tree. These edges are denoted by a \ttt{0} being appended to their label. They need to be distinguished since their green mark needs to be preserved in order for the program to work correctly.

\begin{figure}[p!]
		
\fbox{\begin{minipage}{.98\textwidth}

\begin{allintypewriter}
\smallskip
FindEdge = find\_init; create\_flag; FindLoop!; destroy\_flag

\smallskip
FindLoop = find\_forward!; if flag then Minimise! else 

\phantom{...........}(try MinSetup); try find\_back else break

\smallskip
MinSetup = try min\_init2 then Success else (try min\_init1 then Success)

\smallskip
Success = MinWithS!; set\_flag

\smallskip
Minimise = try MinWithN else MinWithoutN

\smallskip
MinWithS = \string{min\_s, min\_sn, min\_sp, min\_snp, min1\_st, min2\_st\string}

\smallskip
MinWithN = \string{min\_n, min\_np, min\_sn, min\_snp, min\_tn, min\_tnp\string}

\smallskip
MinWithoutN = \string{min, min\_p, min\_s, min\_sp, min\_t, min\_tp, min1\_st, min2\_st\string}

\setlength{\tabcolsep}{5.5pt}
\begin{tabular}{  p{3cm} p{4.2cm} p{3.9cm}  }
\footnotesize

\end{tabular}

\setlength{\tabcolsep}{5.5pt}
\begin{tabular}{  p{3.7cm} p{5cm} p{6cm}  }
			
	find\_init (x:list) & min\_init1 (a,x,y:list) & min\_init2 (i:int; x,y:list) \\ 
	
	\adjustbox{valign=t}{\begin{tikzpicture}
		\node (a) at (0,0) [gp2 root, fill=gp2blue] {x};
		\node (c) at (.65,0)     {$\Rightarrow$};
		\node (d) at (1.3,0)     [gp2 root, fill=gp2grey] {x};
		\node (A) at (0,-.42)   {\tiny{1}};
		\node (D) at (1.3,-.42)   {\tiny{1}};
		\end{tikzpicture}}
	&
	
	\adjustbox{valign=t}{\begin{tikzpicture}
		\node (a) at (0,0) [gp2 root, fill=gp2grey] {x};
		\node (b) at (1,0) [gp2 node, fill=gp2red] {y};
		\node (c) at (1.65,0)     {$\Rightarrow$};
		\node (d) at (2.3,0)     [gp2 root, fill=gp2grey] {x};
		\node (e) at (3.3,0)     [gp2 root, fill=gp2red] {y};
		\node (A) at (0,-.42)   {\tiny{1}};
		\node (B) at (1,-.42)   {\tiny{2}};
		\node (D) at (2.3,-.42)   {\tiny{1}};
		\node (E) at (3.3,-.42)   {\tiny{2}};
		\draw
    		(a) edge[-] node[above] {a} (b)
    		(d) edge[-,gp2green] node[above,black] {a} (e)
		;
		\end{tikzpicture}}
	&
	
	\adjustbox{valign=t}{\begin{tikzpicture}
		\node (a) at (0,0) [gp2 root, fill=gp2grey] {x};
		\node (b) at (1,0) [gp2 node, fill=gp2red] {y};
		\node (c) at (1.65,0)     {$\Rightarrow$};
		\node (d) at (2.3,0)     [gp2 root, fill=gp2grey] {x};
		\node (e) at (3.5,0)     [gp2 root, fill=gp2red] {y};
		\node (A) at (0,-.42)   {\tiny{1}};
		\node (B) at (1,-.42)   {\tiny{2}};
		\node (D) at (2.3,-.42)   {\tiny{1}};
		\node (E) at (3.5,-.42)   {\tiny{2}};
		\draw
    		(a) edge[-,gp2green] node[above,black] {i} (b)
    		(d) edge[-,gp2green] node[above,black] {i:0} (e)
		;
		\end{tikzpicture}}

\normalsize
\end{tabular}

\setlength{\tabcolsep}{5.5pt}
\begin{tabular}{  p{3.3cm} p{2.5cm} p{2.5cm} p{5cm} }
			
	create\_flag () & set\_flag() & flag () & destroy\_flag (x:list) \\ 
	
	\adjustbox{valign=t}{\begin{tikzpicture}
		\node (a) at (0,0) [] {$\emptyset$};
		\node (c) at (.65,0)     {$\Rightarrow$};
		\node (d) at (1.3,0)     [gp2 green root] {0};
		\end{tikzpicture}}
	&
	
	\adjustbox{valign=t}{\begin{tikzpicture}
		\node (a) at (0,0) [gp2 green root] {0};
		\node (c) at (.65,0)     {$\Rightarrow$};
		\node (d) at (1.3,0)     [gp2 green root] {1};
		
		\node (A) at (0,-.42)   {\tiny{1}};
		\node (D) at (1.3,-.42)   {\tiny{1}};
		\end{tikzpicture}}
	&
	
	\adjustbox{valign=t}{\begin{tikzpicture}
		\node (a) at (0,0) [gp2 green root] {1};
		\node (c) at (.65,0)     {$\Rightarrow$};
		\node (d) at (1.3,0)     [gp2 green root] {1};
		
		\node (A) at (0,-.42)   {\tiny{1}};
		\node (D) at (1.3,-.42)   {\tiny{1}};
		\end{tikzpicture}}
	&
	
	\adjustbox{valign=t}{\begin{tikzpicture}
		\node (a) at (0,0) [gp2 green root] {x};
		\node (c) at (.65,0)     {$\Rightarrow$};
		\node (d) at (1.3,0)     [] {$\emptyset$};
		\end{tikzpicture}}

\normalsize
\end{tabular}

\setlength{\tabcolsep}{8pt}
\begin{tabular}{  p{7cm}  p{7cm}  }

	\vspace{-1mm} find\_forward (a,x,y:list) & \vspace{-1mm} find\_back (a,x,y:list)\\
	
	\adjustbox{valign=t}{\begin{tikzpicture}
		\node (a) at (0,0) [gp2 root, fill=gp2grey] {x};
		\node (b) at (1,0) [gp2 node, fill=gp2blue] {y};
		\node (c) at (1.75,0)     {$\Rightarrow$};
		\node (d) at (2.5,0)     [gp2 node, fill=gp2grey] {x};
		\node (e) at (3.5,0)     [gp2 root, fill=gp2grey] {y};
		\node (A) at (0,-.42)   {\tiny{1}};
		\node (B) at (1,-.42)   {\tiny{2}};
		\node (D) at (2.5,-.42)   {\tiny{1}};
		\node (E) at (3.5,-.42)   {\tiny{2}};
		\draw
    		(a) edge[-,performanceBlue] node[above,black] {a} (b)
    		(d) edge[-,dashed] node[above] {a} (e)
		;
		\end{tikzpicture}}
	&		
			
	\adjustbox{valign=t}{\begin{tikzpicture}
		\node (a) at (0,0) [gp2 node, fill=gp2grey] {x};
		\node (b) at (1,0) [gp2 root, fill=gp2grey] {y};
		\node (c) at (1.75,0)     {$\Rightarrow$};
		\node (d) at (2.5,0)     [gp2 root, fill=gp2grey] {x};
		\node (e) at (3.5,0)     [gp2 node, fill=gp2grey] {y};
		\node (A) at (0,-.42)   {\tiny{1}};
		\node (B) at (1,-.42)   {\tiny{2}};
		\node (D) at (2.5,-.42)   {\tiny{1}};
		\node (E) at (3.5,-.42)   {\tiny{2}};
		\draw
    		(d) edge[-,performanceBlue] node[above,black] {a} (e)
    		(a) edge[-,dashed] node[above] {a} (b)
		;
		\end{tikzpicture}}	
		
	\\
	
	\vspace{-1mm} min1\_st (i,j:int; x,y:list) & \vspace{-1mm} min2\_st (i,j:int; x,y:list) \\
	
	\adjustbox{valign=t}{\begin{tikzpicture}
		\node (a) at (0,0) [gp2 root, fill=gp2grey] {x};
		\node (b) at (1,0) [gp2 root, fill=gp2red] {y};
		\node (c) at (1.75,0)     {$\Rightarrow$};
		\node (d) at (2.5,0)     [gp2 root, fill=gp2grey] {x};
		\node (e) at (3.5,0)     [gp2 root, fill=gp2red] {y};
		\node (A) at (0,-.42)   {\tiny{1}};
		\node (B) at (1,-.42)   {\tiny{2}};
		\node (D) at (2.5,-.42)   {\tiny{1}};
		\node (E) at (3.5,-.42)   {\tiny{2}};
		\draw
    		(a) edge[->,gp2green,bend right] node[below,black] {j} (b)
    		(a) edge[-,bend left] node[above,black] {i} (b)
    		(d) edge[-,gp2green,bend left] node[above,black] {i} (e)
    		(d) edge[->,bend right] node[below,black] {j} (e)
		;
		\node[anchor=west] (cond) at (-.35, -.9) {where i<j};
		\end{tikzpicture}}	
	&		
	
	\adjustbox{valign=t}{\begin{tikzpicture}
		\node (a) at (0,0) [gp2 root, fill=gp2grey] {x};
		\node (b) at (1,0) [gp2 root, fill=gp2red] {y};
		\node (c) at (1.75,0)     {$\Rightarrow$};
		\node (d) at (2.5,0)     [gp2 root, fill=gp2grey] {x};
		\node (e) at (3.5,0)     [gp2 root, fill=gp2red] {y};
		\node (A) at (0,-.42)   {\tiny{1}};
		\node (B) at (1,-.42)   {\tiny{2}};
		\node (D) at (2.5,-.42)   {\tiny{1}};
		\node (E) at (3.5,-.42)   {\tiny{2}};
		\draw
    		(a) edge[<-,gp2green,bend right] node[below,black] {j} (b)
    		(a) edge[-,bend left] node[above,black] {i} (b)
    		(d) edge[-,gp2green,bend left] node[above,black] {i} (e)
    		(d) edge[<-,bend right] node[below,black] {j} (e)
		;
		\node[anchor=west] (cond) at (-.35, -.9) {where i<j};
		\end{tikzpicture}}

\normalsize			
\end{tabular}
\end{allintypewriter}

\end{minipage}
}

\caption{The procedure \texttt{FindEdge}}
\label{fig:mst-findedge}
\end{figure}

\begin{figure}[p!]
		
\fbox{\begin{minipage}{.98\textwidth}

\begin{allintypewriter}

\setlength{\tabcolsep}{8pt}
\begin{tabular}{  p{7cm}  p{7cm}  }
	
	\vspace{-1mm} min\_s (i,j:int; x,y,z:list) & \vspace{-1mm} min\_sn (i,j:int; x,y,z:list)\\
	
	\adjustbox{valign=t}{\begin{tikzpicture}
		\node (a) at (0,0) [gp2 root, fill=gp2grey] {x};
		\node (b) at (1,0) [gp2 node, fill=gp2red] {y};
		\node (f) at (1,-.9) [gp2 root, fill=gp2red] {z};
		\node (c) at (1.75,0)     {$\Rightarrow$};
		\node (d) at (2.5,0)     [gp2 root, fill=gp2grey] {x};
		\node (e) at (3.5,0)     [gp2 root, fill=gp2red] {y};
		\node (g) at (3.5,-.9) [gp2 node, fill=gp2red] {z};
		\node (A) at (0,-.42)   {\tiny{1}};
		\node (B) at (1,-.42)   {\tiny{2}};
		\node (F) at (1,-1.32) {\tiny{3}};
		\node (D) at (2.5,-.42)   {\tiny{1}};
		\node (E) at (3.5,-.42)   {\tiny{2}};
		\node (G) at (3.5,-1.32) {\tiny{3}};
		\draw
    		(a) edge[-,gp2green] node[below left,black] {j} (f)
    		(a) edge[-] node[above,black] {i} (b)
    		(d) edge[-,gp2green] node[above,black] {i} (e)
    		(d) edge[-] node[below left,black] {j} (g)
		;
		\node[anchor=west] (cond) at (-.35, -1.7) {where i<j};
		\end{tikzpicture}}	
	&		
	
	\adjustbox{valign=t}{\begin{tikzpicture}
		\node (a) at (0,0) [gp2 root, fill=gp2grey] {x};
		\node (b) at (1,0) [gp2 node, fill=gp2red] {y};
		\node (f) at (1,-.9) [gp2 root, fill=gp2red] {z};
		\node (c) at (1.75,0)     {$\Rightarrow$};
		\node (d) at (2.5,0)     [gp2 root, fill=gp2grey] {x};
		\node (e) at (3.7,0)     [gp2 root, fill=gp2red] {y};
		\node (g) at (3.7,-.9) [gp2 node, fill=gp2red] {z};
		\node (A) at (0,-.42)   {\tiny{1}};
		\node (B) at (1,-.42)   {\tiny{2}};
		\node (F) at (1,-1.32) {\tiny{3}};
		\node (D) at (2.5,-.42)   {\tiny{1}};
		\node (E) at (3.7,-.42)   {\tiny{2}};
		\node (G) at (3.7,-1.32) {\tiny{3}};
		\draw
    		(a) edge[-,gp2green] node[below left,black] {j} (f)
    		(a) edge[-,gp2green] node[above,black] {i} (b)
    		(d) edge[-,gp2green] node[above,black] {i:0} (e)
    		(d) edge[-] node[below left,black] {j} (g)
		;
		\node[anchor=west] (cond) at (-.35, -1.7) {where i<=j};
		\end{tikzpicture}}	
	
	\\
	
	\vspace{-1mm} min\_sp (i,j:int; x,y,z:list) & \vspace{-1mm} min\_snp (i,j:int; x,y,z:list)\\
	
	\adjustbox{valign=t}{\begin{tikzpicture}
		\node (a) at (0,0) [gp2 root, fill=gp2grey] {x};
		\node (b) at (1,0) [gp2 node, fill=gp2red] {y};
		\node (f) at (1,-.9) [gp2 root, fill=gp2red] {z};
		\node (c) at (1.75,0)     {$\Rightarrow$};
		\node (d) at (2.5,0)     [gp2 root, fill=gp2grey] {x};
		\node (e) at (3.5,0)     [gp2 root, fill=gp2red] {y};
		\node (g) at (3.5,-.9) [gp2 node, fill=gp2red] {z};
		\node (A) at (0,-.42)   {\tiny{1}};
		\node (B) at (1,-.42)   {\tiny{2}};
		\node (F) at (1,-1.32) {\tiny{3}};
		\node (D) at (2.5,-.42)   {\tiny{1}};
		\node (E) at (3.5,-.42)   {\tiny{2}};
		\node (G) at (3.5,-1.32) {\tiny{3}};
		\draw
    		(a) edge[-,gp2green] node[below left,black] {j:0} (f)
    		(a) edge[-] node[above,black] {i} (b)
    		(d) edge[-,gp2green] node[above,black] {i} (e)
    		(d) edge[-,gp2green] node[below left,black] {j} (g)
		;
		\node[anchor=west] (cond) at (-.35, -1.7) {where i<j};
		\end{tikzpicture}}	
    
	&		
	
	\adjustbox{valign=t}{\begin{tikzpicture}
		\node (a) at (0,0) [gp2 root, fill=gp2grey] {x};
		\node (b) at (1,0) [gp2 node, fill=gp2red] {y};
		\node (f) at (1,-.9) [gp2 root, fill=gp2red] {z};
		\node (c) at (1.75,0)     {$\Rightarrow$};
		\node (d) at (2.5,0)     [gp2 root, fill=gp2grey] {x};
		\node (e) at (3.7,0)     [gp2 root, fill=gp2red] {y};
		\node (g) at (3.7,-.9) [gp2 node, fill=gp2red] {z};
		\node (A) at (0,-.42)   {\tiny{1}};
		\node (B) at (1,-.42)   {\tiny{2}};
		\node (F) at (1,-1.32) {\tiny{3}};
		\node (D) at (2.5,-.42)   {\tiny{1}};
		\node (E) at (3.7,-.42)   {\tiny{2}};
		\node (G) at (3.7,-1.32) {\tiny{3}};
		\draw
    		(a) edge[-,gp2green] node[below left,black] {j:0} (f)
    		(a) edge[-,gp2green] node[above,black] {i} (b)
    		(d) edge[-,gp2green] node[above,black] {i:0} (e)
    		(d) edge[-,gp2green] node[below left,black] {j} (g)
		;
		\node[anchor=west] (cond) at (-.35, -1.7) {where i<j};
		\end{tikzpicture}}	
		
		\\
		
		\vspace{-1mm} min\_t (i,j:int; x,y,z:list) & \vspace{-1mm} min\_tn (i,j:int; x,y,z:list)\\
	
	\adjustbox{valign=t}{\begin{tikzpicture}
		\node (a) at (0,0) [gp2 root, fill=gp2grey] {x};
		\node (b) at (1,0) [gp2 root, fill=gp2red] {y};
		\node (h) at (0,-.9) [gp2 node, fill=gp2grey] {z};
		\node (c) at (1.75,0)     {$\Rightarrow$};
		\node (d) at (2.5,0)     [gp2 root, fill=gp2grey] {x};
		\node (e) at (3.5,0)     [gp2 root, fill=gp2red] {y};
		\node (i) at (2.5,-.9) [gp2 node, fill=gp2grey] {z};
		\node (A) at (0,-.42)   {\tiny{1}};
		\node (B) at (1,-.42)   {\tiny{2}};
		\node (H) at (0,-1.32) {\tiny{3}};
		\node (D) at (2.5,-.42)   {\tiny{1}};
		\node (E) at (3.5,-.42)   {\tiny{2}};
		\node (I) at (2.5,-1.32) {\tiny{3}};
		\draw
    		(h) edge[-,gp2green] node[below right,black] {j} (b)
    		(a) edge[-] node[above,black] {i} (b)
    		(d) edge[-,gp2green] node[above,black] {i} (e)
    		(i) edge[-] node[below right,black] {j} (e)
		;
		\node[anchor=west] (cond) at (-.35, -1.7) {where i<j};
		\end{tikzpicture}}
    	
	&		
	
	\adjustbox{valign=t}{\begin{tikzpicture}
		\node (a) at (0,0) [gp2 root, fill=gp2grey] {x};
		\node (b) at (1,0) [gp2 root, fill=gp2red] {y};
		\node (h) at (0,-.9) [gp2 node, fill=gp2grey] {z};
		\node (c) at (1.75,0)     {$\Rightarrow$};
		\node (d) at (2.5,0)     [gp2 root, fill=gp2grey] {x};
		\node (e) at (3.7,0)     [gp2 root, fill=gp2red] {y};
		\node (i) at (2.5,-.9) [gp2 node, fill=gp2grey] {z};
		\node (A) at (0,-.42)   {\tiny{1}};
		\node (B) at (1,-.42)   {\tiny{2}};
		\node (H) at (0,-1.32) {\tiny{3}};
		\node (D) at (2.5,-.42)   {\tiny{1}};
		\node (E) at (3.7,-.42)   {\tiny{2}};
		\node (I) at (2.5,-1.32) {\tiny{3}};
		\draw
    		(h) edge[-,gp2green] node[below right,black] {j} (b)
    		(a) edge[-,gp2green] node[above,black] {i} (b)
    		(d) edge[-,gp2green] node[above,black] {i:0} (e)
    		(i) edge[-] node[below right,black] {j} (e)
		;
		\node[anchor=west] (cond) at (-.35, -1.7) {where i<=j};
		\end{tikzpicture}}
	
	\\
	
	\vspace{-1mm} min\_tp (i,j:int; x,y,z:list) & \vspace{-1mm} min\_tnp (i,j:int; x,y,z:list)\\
	
	\adjustbox{valign=t}{\begin{tikzpicture}
		\node (a) at (0,0) [gp2 root, fill=gp2grey] {x};
		\node (b) at (1.2,0) [gp2 root, fill=gp2red] {y};
		\node (h) at (0,-.9) [gp2 node, fill=gp2grey] {z};
		\node (c) at (1.95,0)     {$\Rightarrow$};
		\node (d) at (2.7,0)     [gp2 root, fill=gp2grey] {x};
		\node (e) at (3.7,0)     [gp2 root, fill=gp2red] {y};
		\node (i) at (2.7,-.9) [gp2 node, fill=gp2grey] {z};
		\node (A) at (0,-.42)   {\tiny{1}};
		\node (B) at (1.2,-.42)   {\tiny{2}};
		\node (H) at (0,-1.32) {\tiny{3}};
		\node (D) at (2.7,-.42)   {\tiny{1}};
		\node (E) at (3.7,-.42)   {\tiny{2}};
		\node (I) at (2.7,-1.32) {\tiny{3}};
		\draw
    		(h) edge[-,gp2green] node[below right,black] {j:0} (b)
    		(a) edge[-] node[above,black] {i} (b)
    		(d) edge[-,gp2green] node[above,black] {i} (e)
    		(i) edge[-,gp2green] node[below right,black] {j} (e)
		;
		\node[anchor=west] (cond) at (-.35, -1.7) {where i<j};
		\end{tikzpicture}}
    
	&		
	
	\adjustbox{valign=t}{\begin{tikzpicture}
		\node (a) at (0,0) [gp2 root, fill=gp2grey] {x};
		\node (b) at (1.2,0) [gp2 node, fill=gp2red] {y};
		\node (h) at (0,-.9) [gp2 node, fill=gp2grey] {z};
		\node (c) at (1.95,0)     {$\Rightarrow$};
		\node (d) at (2.7,0)     [gp2 root, fill=gp2grey] {x};
		\node (e) at (3.9,0)     [gp2 root, fill=gp2red] {y};
		\node (i) at (2.7,-.9) [gp2 node, fill=gp2grey] {z};
		\node (A) at (0,-.42)   {\tiny{1}};
		\node (B) at (1.2,-.42)   {\tiny{2}};
		\node (H) at (0,-1.32) {\tiny{3}};
		\node (D) at (2.7,-.42)   {\tiny{1}};
		\node (E) at (3.9,-.42)   {\tiny{2}};
		\node (I) at (2.7,-1.32) {\tiny{3}};
		\draw
    		(h) edge[-,gp2green] node[below right,black] {j:0} (b)
    		(a) edge[-,gp2green] node[above,black] {i} (b)
    		(d) edge[-,gp2green] node[above,black] {i:0} (e)
    		(i) edge[-,gp2green] node[below right,black] {j} (e)
		;
		\node[anchor=west] (cond) at (-.35, -1.7) {where i<j};
		\end{tikzpicture}}
		
	\\
	
	\vspace{-1mm} min (i,j:int; x,y,z,t:list) & \vspace{-1mm} min\_n (i,j:int; x,y,z,t:list)\\
	
	\adjustbox{valign=t}{\begin{tikzpicture}
		\node (a) at (0,0) [gp2 root, fill=gp2grey] {x};
		\node (b) at (1,0) [gp2 node, fill=gp2red] {y};
		\node (h) at (0,-.9) [gp2 node, fill=gp2grey] {z};
		\node (f) at (1,-.9) [gp2 root, fill=gp2red] {t};
		\node (c) at (1.75,0)     {$\Rightarrow$};
		\node (d) at (2.5,0)     [gp2 root, fill=gp2grey] {x};
		\node (e) at (3.5,0)     [gp2 root, fill=gp2red] {y};
		\node (i) at (2.5,-.9) [gp2 node, fill=gp2grey] {z};
		\node (g) at (3.5,-.9) [gp2 node, fill=gp2red] {t};
		\node (A) at (0,-.42)   {\tiny{1}};
		\node (B) at (1,-.42)   {\tiny{2}};
		\node (H) at (0,-1.32) {\tiny{3}};
		\node (F) at (1,-1.32) {\tiny{4}};
		\node (D) at (2.5,-.42)   {\tiny{1}};
		\node (E) at (3.5,-.42)   {\tiny{2}};
		\node (I) at (2.5,-1.32) {\tiny{3}};
		\node (G) at (3.5,-1.32) {\tiny{4}};
		\draw
    		(h) edge[-,gp2green] node[above,black] {j} (f)
    		(a) edge[-] node[above,black] {i} (b)
    		(d) edge[-,gp2green] node[above,black] {i} (e)
    		(i) edge[-] node[above,black] {j} (g)
		;
		\node[anchor=west] (cond) at (-.35, -1.7) {where i<j};
		\end{tikzpicture}}
    	
	&		
	
	\adjustbox{valign=t}{\begin{tikzpicture}
		\node (a) at (0,0) [gp2 root, fill=gp2grey] {x};
		\node (b) at (1,0) [gp2 node, fill=gp2red] {y};
		\node (h) at (0,-.9) [gp2 node, fill=gp2grey] {z};
		\node (f) at (1,-.9) [gp2 root, fill=gp2red] {t};
		\node (c) at (1.75,0)     {$\Rightarrow$};
		\node (d) at (2.5,0)     [gp2 root, fill=gp2grey] {x};
		\node (e) at (3.7,0)     [gp2 root, fill=gp2red] {y};
		\node (i) at (2.5,-.9) [gp2 node, fill=gp2grey] {z};
		\node (g) at (3.7,-.9) [gp2 node, fill=gp2red] {t};
		\node (A) at (0,-.42)   {\tiny{1}};
		\node (B) at (1,-.42)   {\tiny{2}};
		\node (H) at (0,-1.32) {\tiny{3}};
		\node (F) at (1,-1.32) {\tiny{4}};
		\node (D) at (2.5,-.42)   {\tiny{1}};
		\node (E) at (3.7,-.42)   {\tiny{2}};
		\node (I) at (2.5,-1.32) {\tiny{3}};
		\node (G) at (3.7,-1.32) {\tiny{4}};
		\draw
    		(h) edge[-,gp2green] node[above,black] {j} (f)
    		(a) edge[-,gp2green] node[above,black] {i} (b)
    		(d) edge[-,gp2green] node[above,black] {i:0} (e)
    		(i) edge[-] node[above,black] {j} (g)
		;
		\node[anchor=west] (cond) at (-.35, -1.7) {where i<=j};
		\end{tikzpicture}}
	
	\\
	
	\vspace{-1mm} min\_p (i,j:int; x,y,z,t:list) & \vspace{-1mm} min\_np (i,j:int; x,y,z,t:list)\\
	
	\adjustbox{valign=t}{\begin{tikzpicture}
		\node (a) at (0,0) [gp2 root, fill=gp2grey] {x};
		\node (b) at (1.2,0) [gp2 node, fill=gp2red] {y};
		\node (h) at (0,-.9) [gp2 node, fill=gp2grey] {z};
		\node (f) at (1.2,-.9) [gp2 root, fill=gp2red] {t};
		\node (c) at (1.95,0)     {$\Rightarrow$};
		\node (d) at (2.7,0)     [gp2 root, fill=gp2grey] {x};
		\node (e) at (3.7,0)     [gp2 root, fill=gp2red] {y};
		\node (i) at (2.7,-.9) [gp2 node, fill=gp2grey] {z};
		\node (g) at (3.7,-.9) [gp2 node, fill=gp2red] {t};
		\node (A) at (0,-.42)   {\tiny{1}};
		\node (B) at (1.2,-.42)   {\tiny{2}};
		\node (H) at (0,-1.32) {\tiny{3}};
		\node (F) at (1.2,-1.32) {\tiny{4}};
		\node (D) at (2.7,-.42)   {\tiny{1}};
		\node (E) at (3.7,-.42)   {\tiny{2}};
		\node (I) at (2.7,-1.32) {\tiny{3}};
		\node (G) at (3.7,-1.32) {\tiny{4}};
		\draw
    		(h) edge[-,gp2green] node[above,black] {j:0} (f)
    		(a) edge[-] node[above,black] {i} (b)
    		(d) edge[-,gp2green] node[above,black] {i} (e)
    		(i) edge[-,gp2green] node[above,black] {j} (g)
		;
		\node[anchor=west] (cond) at (-.35, -1.7) {where i<j};
		\end{tikzpicture}}
    
	&		
	
	\adjustbox{valign=t}{\begin{tikzpicture}
		\node (a) at (0,0) [gp2 root, fill=gp2grey] {x};
		\node (b) at (1.2,0) [gp2 node, fill=gp2red] {y};
		\node (h) at (0,-.9) [gp2 node, fill=gp2grey] {z};
		\node (f) at (1.2,-.9) [gp2 root, fill=gp2red] {t};
		\node (c) at (1.95,0)     {$\Rightarrow$};
		\node (d) at (2.7,0)     [gp2 root, fill=gp2grey] {x};
		\node (e) at (3.9,0)     [gp2 root, fill=gp2red] {y};
		\node (i) at (2.7,-.9) [gp2 node, fill=gp2grey] {z};
		\node (g) at (3.9,-.9) [gp2 node, fill=gp2red] {t};
		\node (A) at (0,-.42)   {\tiny{1}};
		\node (B) at (1.2,-.42)   {\tiny{2}};
		\node (H) at (0,-1.32) {\tiny{3}};
		\node (F) at (1.2,-1.32) {\tiny{4}};
		\node (D) at (2.7,-.42)   {\tiny{1}};
		\node (E) at (3.9,-.42)   {\tiny{2}};
		\node (I) at (2.7,-1.32) {\tiny{3}};
		\node (G) at (3.9,-1.32) {\tiny{4}};
		\draw
    		(h) edge[-,gp2green] node[above,black] {j:0} (f)
    		(a) edge[-,gp2green] node[above,black] {i} (b)
    		(d) edge[-,gp2green] node[above,black] {i:0} (e)
    		(i) edge[-,gp2green] node[above,black] {j} (g)
		;
		\node[anchor=west] (cond) at (-.35, -1.7) {where i<j};
		\end{tikzpicture}}
	
\normalsize			
\end{tabular}
\end{allintypewriter}

\end{minipage}
}

\caption{More ``\texttt{min}'' rules}
\label{fig:mst-min}
\end{figure}

The ``min'' rules with both the \ttt{s} and \ttt{t} flags, i.e. the ones minimising parallel edges, are a special case. We omit the cases that involve previously selected edges (flags \ttt{n} or \ttt{p}) since such an edge would have already been minimised over its parallel edges by previous applications of \ttt{min1\_st} and \ttt{min2\_st}. We use two rules with directed edges labelled \ttt{j} instead of one rule with a bidirectional edge labelled \ttt{j} due to parallel bidirectional edges being disallowed by GP\texorpdfstring{\,}{ }2. This is because, if the parallel bidirectional edges are indistinguishable in the left hand side of a rule, the result of the rule application is not necessarily unique up to isomorphism, since it could leave the host graph with an edge in one of two possible directions.

In order to prioritise edges that have already been selected for different trees, we call the rules of the procedure \ttt{MinWithN} first. They consist of the rules with flag \ttt{n}. We can then call the rest of the rules using \ttt{MinWithoutN}.
Note that \ttt{min\_sn}, \ttt{min\_tn}, and \ttt{min\_n} (i.e. ``min'' rules with \ttt{n} but not \ttt{p}) are the only rules that can be applied if the weights are equal. This is because they are the only ones selecting a previously selected edge, which we prioritise. Making the other rules applicable on equal weights can lead to non-termination. In our example, \ttt{min} is applied.

Finally, the DFS terminates and the rule \ttt{destroy\_flag} deletes the temporary flag needed for this procedure. The flag could have been implemented as an additional list entry of the unmarked root, but was chosen to be its own green root for the sake of semantic clarity.

\subsection{The Procedure \ttt{GrowForest}}
\label{subsec:growforest}

The procedure \ttt{GrowForest} depicted in Figure \ref{fig:mst-GrowForest} serves to turn the edges selected by \ttt{FindEdge!} (green mark) into edges of the forest (blue mark), thus merging some of the trees. Graphs 6 and 7 in the example execution of \ttt{mst-boruvka} in Figure \ref{fig:ex-grid} exemplify input and output graph of \ttt{GrowForest}.

\begin{figure}[ht]
		
\fbox{\begin{minipage}{.98\textwidth}

\begin{allintypewriter}
GrowForest = grow\_init; GrowLoop!; GrowClean!; unroot\_red

\smallskip
GrowLoop = GrowTree! try next\_root else break

\smallskip
GrowTree = down!; add\_edge!; try up else break

\smallskip
GrowClean = try ColourRed; try previous\_root else break

\medskip
\setlength{\tabcolsep}{8pt}
\begin{tabular}{  p{7cm}  p{7cm}  }
			
	grow\_init (x,y:list) & down (a,x,y:list) \\ 
	
	\adjustbox{valign=t}{\begin{tikzpicture}
		\node (a) at (0,0)     [gp2 root] {x};
		\node (b) at (1,0)     [gp2 red node] {y};
		\node (c) at (1.75,0)     {$\Rightarrow$};
		\node (d) at (2.5,0)     [gp2 root] {x};
		\node (e) at (3.5,0)     [gp2 grey root] {y};
		\node (A) at (0,-.42)   {\tiny{1}};
		\node (B) at (1,-.42)   {\tiny{2}};
		\node (D) at (2.5,-.42)   {\tiny{1}};
		\node (E) at (3.5,-.42)   {\tiny{2}};
		\draw
		(a) edge[->,gp2red] (b)
		(d) edge[->,gp2red] (e)
		;
		\end{tikzpicture}}
	&
	
	\adjustbox{valign=t}{\begin{tikzpicture}
		\node (a) at (0,0)     [gp2 grey root] {x};
		\node (b) at (1,0)     [gp2 red node] {y};
		\node (c) at (1.75,0)     {$\Rightarrow$};
		\node (d) at (2.5,0)     [gp2 grey node] {x};
		\node (e) at (3.5,0)     [gp2 grey root] {y};
		\node (A) at (0,-.42)   {\tiny{1}};
		\node (B) at (1,-.42)   {\tiny{2}};
		\node (D) at (2.5,-.42)   {\tiny{1}};
		\node (E) at (3.5,-.42)   {\tiny{2}};
		\draw (a) edge[-,performanceBlue] node[above,black] {a} (b)
		(d) edge[-,dashed] node[above] {a} (e)
		;
		\end{tikzpicture}}
	\\

	\vspace{-1mm} add\_edge (a,x,y:list) & \vspace{-1mm} up (a,x,y:list) \\  
	
	\adjustbox{valign=t}{\begin{tikzpicture}
		\node (a) at (0,0) [gp2 grey root] {x};
		\node (b) at (1,0) [gp2 red node] {y};
		\node (c) at (1.75,0)     {$\Rightarrow$};
		\node (d) at (2.5,0)     [gp2 grey root] {x};
		\node (e) at (3.5,0)     [gp2 red node] {y};
		\node (A) at (0,-.42)   {\tiny{1}};
		\node (B) at (1,-.42)   {\tiny{2}};
		\node (D) at (2.5,-.42)   {\tiny{1}};
		\node (E) at (3.5,-.42)   {\tiny{2}};
		\draw
    		(a) edge[-,gp2green] node[above,black] {a} (b)
    		(d) edge[-,performanceBlue] node[above,black] {a} (e)
		;
		\end{tikzpicture}}
	
	& 
	
	\adjustbox{valign=t}{\begin{tikzpicture}
		\node (a) at (0,0)     [gp2 grey node] {x};
		\node (b) at (1,0)     [gp2 grey root] {y};
		\node (c) at (1.75,0)     {$\Rightarrow$};
		\node (d) at (2.5,0)     [gp2 grey root] {x};
		\node (e) at (3.5,0)     [gp2 grey node] {y};
		\node (A) at (0,-.42)   {\tiny{1}};
		\node (B) at (1,-.42)   {\tiny{2}};
		\node (D) at (2.5,-.42)   {\tiny{1}};
		\node (E) at (3.5,-.42)   {\tiny{2}};
		\draw
		(a) edge[-,dashed] node[above] {a} (b)
		(d) edge[-,performanceBlue] node[above,black] {a} (e)
		;
		\end{tikzpicture}}
	\\
	
	\vspace{-1mm} next\_root (x,y:list) & \vspace{-1mm} previous\_root (x,y:list) \\  
	
	\adjustbox{valign=t}{\begin{tikzpicture}
		\node (a) at (0,0) [gp2 red node] {x};
		\node (b) at (1,0) [gp2 grey root] {y};
		\node (c) at (1.75,0)     {$\Rightarrow$};
		\node (d) at (2.5,0)     [gp2 grey root] {x};
		\node (e) at (3.5,0)     [gp2 grey node] {y};
		\node (A) at (0,-.42)   {\tiny{1}};
		\node (B) at (1,-.42)   {\tiny{2}};
		\node (D) at (2.5,-.42)   {\tiny{1}};
		\node (E) at (3.5,-.42)   {\tiny{2}};
		\draw
    		(a) edge[->,gp2red] (b)
    		(d) edge[->,gp2red] (e)
		;
		\end{tikzpicture}}
	
	& 
	
	\adjustbox{valign=t}{\begin{tikzpicture}
		\node (a) at (0,0)     [gp2 red root] {x};
		\node (b) at (1,0)     [gp2 any node] {y};
		\node (c) at (1.75,0)     {$\Rightarrow$};
		\node (d) at (2.5,0)     [gp2 red node] {x};
		\node (e) at (3.5,0)     [gp2 any root] {y};
		\node (A) at (0,-.42)   {\tiny{1}};
		\node (B) at (1,-.42)   {\tiny{2}};
		\node (D) at (2.5,-.42)   {\tiny{1}};
		\node (E) at (3.5,-.42)   {\tiny{2}};
		\draw
		(a) edge[->,gp2red] (b)
		(d) edge[->,gp2red] (e)
		;
		\end{tikzpicture}}
	
\normalsize			
\end{tabular}
\end{allintypewriter}

\end{minipage}
}

\caption{The procedure \texttt{GrowForest}}
\label{fig:mst-GrowForest}
\end{figure}

The procedure \ttt{GrowForest} traverses the graph by iterating through the list of trees, and conducting a DFS on each tree. \ttt{next\_root} helps iterate through the list in the direction opposite to the orientation of the red edges.

The rules \ttt{down} and \ttt{up} play the roles of \ttt{forward} and \ttt{back} in a DFS. They use blue edges to ensure only the current tree is traversed. \ttt{add\_edge!} is called right before \ttt{up} to turn all green edges adjacent to the grey root blue. After the \ttt{up} rule is applied to a grey root, it is not visited again by the DFS, ensuring the new blue edges will not be traversed. Future DFSs will also not traverse these edges since one of its adjacent nodes is grey.

The loop \ttt{CleanUp!} iterates through the list of trees in the direction opposite to \ttt{GrowTree!} and calls \ttt{ColourRed} on each tree to mark the nodes red again.

\subsection{Other Procedures}
\label{subsec:other}

The program \ttt{mst-boruvka} calls several procedures to maintain the list data structure or to prepare the graph for the next step. Now we describe these procedures, \ttt{ColourBlue} in Figure \ref{fig:mst-blue}, \ttt{ColourRed} in Figure \ref{fig:mst-red}, \ttt{MarkForDeletion} in Figure \ref{fig:mst-mark}, and \ttt{Rewind} in Figure \ref{fig:mst-rewind}.


\begin{figure}[ht]
		
\fbox{\begin{minipage}{.98\textwidth}

\begin{allintypewriter}

ColourBlue = blue\_init; BlueLoop!

\smallskip
BlueLoop = blue\_forward!; try blue\_back else break

\medskip
\setlength{\tabcolsep}{5.5pt}
\begin{tabular}{  p{4cm} p{5.6cm} p{6cm}  }
			
	blue\_init (x:list)	& blue\_forward (a,x,y:list) & blue\_back (a,x,y:list)\\ 
	
	\adjustbox{valign=t}{\begin{tikzpicture}
		\node (a) at (0,0)   [gp2 root, fill=gp2red] {x};
		
		\node (b) at (.76,0) {$\Rightarrow$};
		
		\node (c) at (1.3,0) [gp2 root, fill=gp2blue] {x};
		
		\node (A) at (0,-.42)   {\tiny{1}};
		\node (C) at (1.3,-.42) {\tiny{1}};
		\end{tikzpicture}}
	&
	
	\adjustbox{valign=t}{\begin{tikzpicture}
		\node (a) at (0,0) [gp2 root, fill=gp2blue] {x};
		\node (b) at (1,0) [gp2 node, fill=gp2red] {y};
		\node (c) at (1.65,0)     {$\Rightarrow$};
		\node (d) at (2.3,0)     [gp2 node, fill=gp2blue] {x};
		\node (e) at (3.3,0)     [gp2 root, fill=gp2blue] {y};
		\node (A) at (0,-.42)   {\tiny{1}};
		\node (B) at (1,-.42)   {\tiny{2}};
		\node (D) at (2.3,-.42)   {\tiny{1}};
		\node (E) at (3.3,-.42)   {\tiny{2}};
		\draw
    		(a) edge[-,performanceBlue] node[above,black] {a} (b)
    		(d) edge[-,dashed] node[above] {a} (e)
		;
		\end{tikzpicture}}
	&
	
	\adjustbox{valign=t}{\begin{tikzpicture}
		\node (a) at (0,0) [gp2 node, fill=gp2blue] {x};
		\node (b) at (1,0) [gp2 root, fill=gp2blue] {y};
		\node (c) at (1.65,0)     {$\Rightarrow$};
		\node (d) at (2.3,0)     [gp2 root, fill=gp2blue] {x};
		\node (e) at (3.3,0)     [gp2 node, fill=gp2blue] {y};
		\node (A) at (0,-.42)   {\tiny{1}};
		\node (B) at (1,-.42)   {\tiny{2}};
		\node (D) at (2.3,-.42)   {\tiny{1}};
		\node (E) at (3.3,-.42)   {\tiny{2}};
		\draw
    		(a) edge[-,dashed] node[above] {a} (b)
    		(d) edge[-,performanceBlue] node[above,black] {a} (e)
		;
		\end{tikzpicture}}

\normalsize
\end{tabular}
\end{allintypewriter}

\end{minipage}
}

\caption{The procedure \texttt{ColourBlue}}
\label{fig:mst-blue}
\end{figure}


\begin{figure}[ht]
		
\fbox{\begin{minipage}{.98\textwidth}

\begin{allintypewriter}

ColourRed = red\_init; RedLoop!

\smallskip
RedLoop = red\_forward!; try red\_back else break

\medskip
\setlength{\tabcolsep}{5.5pt}
\begin{tabular}{  p{4cm} p{5.6cm} p{6cm}  }
			
	red\_init (x:list)	& red\_forward (a,x,y:list) & red\_back (a,x,y:list)\\ 
	
	\adjustbox{valign=t}{\begin{tikzpicture}
		\node (a) at (0,0)   [gp2 root, fill=gp2grey] {x};
		
		\node (b) at (.65,0) {$\Rightarrow$};
		
		\node (c) at (1.3,0) [gp2 root, fill=gp2red] {x};
		
		\node (A) at (0,-.42)   {\tiny{1}};
		\node (C) at (1.3,-.42) {\tiny{1}};
		\end{tikzpicture}}
	&
	
	\adjustbox{valign=t}{\begin{tikzpicture}
		\node (a) at (0,0) [gp2 root, fill=gp2red] {x};
		\node (b) at (1,0) [gp2 node, fill=gp2grey] {y};
		\node (c) at (1.65,0)     {$\Rightarrow$};
		\node (d) at (2.3,0)     [gp2 node, fill=gp2red] {x};
		\node (e) at (3.3,0)     [gp2 root, fill=gp2red] {y};
		\node (A) at (0,-.42)   {\tiny{1}};
		\node (B) at (1,-.42)   {\tiny{2}};
		\node (D) at (2.3,-.42)   {\tiny{1}};
		\node (E) at (3.3,-.42)   {\tiny{2}};
		\draw
    		(a) edge[-,performanceBlue] node[above,black] {a} (b)
    		(d) edge[-,dashed] node[above] {a} (e)
		;
		\end{tikzpicture}}
	&
	
	\adjustbox{valign=t}{\begin{tikzpicture}
		\node (a) at (0,0) [gp2 node, fill=gp2red] {x};
		\node (b) at (1,0) [gp2 root, fill=gp2red] {y};
		\node (c) at (1.65,0)     {$\Rightarrow$};
		\node (d) at (2.3,0)     [gp2 root, fill=gp2red] {x};
		\node (e) at (3.3,0)     [gp2 node, fill=gp2red] {y};
		\node (A) at (0,-.42)   {\tiny{1}};
		\node (B) at (1,-.42)   {\tiny{2}};
		\node (D) at (2.3,-.42)   {\tiny{1}};
		\node (E) at (3.3,-.42)   {\tiny{2}};
		\draw
    		(a) edge[-,dashed] node[above] {a} (b)
    		(d) edge[-,performanceBlue] node[above,black] {a} (e)
		;
		\end{tikzpicture}}

\normalsize
\end{tabular}
\end{allintypewriter}

\end{minipage}
}

\caption{The procedure \texttt{ColourRed}}
\label{fig:mst-red}
\end{figure}


The procedure \ttt{ColourBlue} uses DFS to turn the nodes of a tree from red to blue, and the procedure \ttt{ColourRed} to turn the nodes of a tree from grey to red.


\begin{figure}[ht]
		
\fbox{\begin{minipage}{.98\textwidth}

\begin{allintypewriter}
MarkForDeletion = try clean else Mark; unroot\_red

\smallskip
Mark = if red\_loop then skip else add\_loop

\medskip
\setlength{\tabcolsep}{8pt}
\begin{tabular}{  p{5cm}  p{3.7cm} p{3.5cm}  }
			
	clean (i:int; x,y:list) & red\_loop (x:list) & add\_loop (x:list) \\ 
	
	\adjustbox{valign=t}{\begin{tikzpicture}
		\node (a) at (0,0) [gp2 grey node] {x};
		\node (b) at (1.2,0) [gp2 red root] {y};
		\node (c) at (1.85,0)     {$\Rightarrow$};
		\node (d) at (2.5,0)     [gp2 grey node] {x};
		\node (e) at (3.5,0)     [gp2 red root] {y};
		\node (A) at (0,-.42)   {\tiny{1}};
		\node (B) at (1.2,-.42)   {\tiny{2}};
		\node (D) at (2.5,-.42)   {\tiny{1}};
		\node (E) at (3.5,-.42)   {\tiny{2}};
		\draw
    		(a) edge[-,gp2green] node[above,black] {i:0} (b)
    		(d) edge[-,gp2green] node[above,black] {i} (e)
		;
		\end{tikzpicture}}
	&
	
	\adjustbox{valign=t}{\begin{tikzpicture}
		\node (a) at (0,0)   [gp2 grey root] {x};
		
		\node (b) at (.75,0) {$\Rightarrow$};
		
		\node (c) at (1.5,0) [gp2 grey root] {x};
		\draw
    		(a) edge[-,gp2red,in=-30,out=-60,loop,thick] (a)
    		(c) edge[-,gp2red,in=-30,out=-60,loop,thick] (c)
		;
		\node (A) at (0,-.42)   {\tiny{1}};
		\node (C) at (1.5,-.42) {\tiny{1}};
	\end{tikzpicture}}
	&
	
	\adjustbox{valign=t}{\begin{tikzpicture}
		\node (a) at (0,0)   [gp2 grey root] {x};
		
		\node (b) at (.75,0) {$\Rightarrow$};
		
		\node (c) at (1.5,0) [gp2 grey root] {x};
		\draw
    		(c) edge[-,gp2red,in=-30,out=-60,loop,thick] (c)
		;
		\node (A) at (0,-.42)   {\tiny{1}};
		\node (C) at (1.5,-.42) {\tiny{1}};
	\end{tikzpicture}}

\normalsize			
\end{tabular}
\end{allintypewriter}

\end{minipage}
}

\caption{The procedure \texttt{MarkForDeletion}}
\label{fig:mst-mark}
\end{figure}

\begin{figure}[ht]
		
\fbox{\begin{minipage}{.98\textwidth}

\begin{allintypewriter}
Rewind = try remove\_mid else RemoveEnd

\smallskip
RemoveEnd = try \string{remove\_top, remove\_bottom\string} else keep

\medskip
\setlength{\tabcolsep}{8pt}
\begin{tabular}{  p{7cm}  p{7cm}  }
			
	remove\_mid (i:int; x,y,z:list) & remove\_top (i:int; x,y:list) \\ 
	
	\adjustbox{valign=t}{\begin{tikzpicture}
	    \node (a) at (0,0) [gp2 red node] {x};
		\node (b) at (1,0) [gp2 red node] {y};
		\node (c) at (1.5,-.9) [gp2 red node] {z};
		\node (d) at (.5,-.9) [gp2 root] {i};
		\draw
    		(a) edge[->, gp2red] (b)
    		(b) edge[->, gp2red] (c)
    		(d) edge[->, gp2red] (b)
    		(b) edge[-,gp2red,in=15,out=-15,loop,thick] (b)
		;
		
		\node (A) at (0,-.42)   {\tiny{1}};
		\node (B) at (1,-.42)   {\tiny{2}};
		\node (C) at (1.5,-1.32)   {\tiny{4}};
		\node (D) at (.5, -1.32)   {\tiny{3}};
		
		\node (c) at (2.25,-.5)     {$\Rightarrow$};
		
		\node (a) at (3,0) [gp2 red node] {x};
		\node (b) at (4,0) [gp2 red node] {y};
		\node (c) at (4.5,-.9) [gp2 red node] {z};
		\node (d) at (3.5,-.9) [gp2 root] {i-1};
		\draw
    		(a) edge[->, gp2red] (c)
    		(d) edge[->, gp2red] (a)
		;
		
		\node (A) at (3,-.42)   {\tiny{1}};
		\node (B) at (4,-.42)   {\tiny{2}};
		\node (C) at (4.5,-1.32)   {\tiny{4}};
		\node (D) at (3.5, -1.32)   {\tiny{3}};
		
		\end{tikzpicture}}
	&
	
	\adjustbox{valign=t}{\begin{tikzpicture}
		\node (a) at (0,0) [gp2 red node] {x};
		\node (b) at (1,0) [gp2 red node] {y};
		\node (c) at (.5,-.9)     [gp2 root] {i};
		\draw
		    (a) edge[->, gp2red] (b)
		    (c) edge[->, gp2red] (a)
		    (a) edge[-,gp2red,in=-20,out=-45,loop,thick] (a)
		;
		\node (A) at (0,-.42)   {\tiny{1}};
		\node (B) at (1,-.42)   {\tiny{2}};
		\node (C) at (.5, -1.32)   {\tiny{3}};
		
		\node (c) at (1.75,-.5)     {$\Rightarrow$};
		
		\node (a) at (2.5,0) [gp2 red node] {x};
		\node (b) at (3.5,0) [gp2 red node] {y};
		\node (c) at (3,-.9)     [gp2 root] {i-1};
		\draw
		    (c) edge[->, gp2red] (b)
		;
		\node (A) at (2.5,-.42)   {\tiny{1}};
		\node (B) at (3.5,-.42)   {\tiny{2}};
		\node (C) at (3, -1.32)   {\tiny{3}};
		\end{tikzpicture}}
	\\

	\vspace{-1mm} remove\_bottom (i:int; x,y:list) & \vspace{-1mm} keep (i:int; x,y:list) \\  
	
	\adjustbox{valign=t}{\begin{tikzpicture}
		\node (a) at (0,0) [gp2 red node] {x};
		\node (b) at (1,0) [gp2 red node] {y};
		\node (c) at (.5,-.9)     [gp2 root] {i};
		\draw
		    (a) edge[->, gp2red] (b)
		    (c) edge[->, gp2red] (b)
		    (b) edge[-,gp2red,in=-160,out=-135,loop,thick] (b)
		;
		\node (A) at (0,-.42)   {\tiny{1}};
		\node (B) at (1,-.42)   {\tiny{2}};
		\node (C) at (.5, -1.32)   {\tiny{3}};
		
		\node (c) at (1.75,-.5)     {$\Rightarrow$};
		
		\node (a) at (2.5,0) [gp2 red node] {x};
		\node (b) at (3.5,0) [gp2 red node] {y};
		\node (c) at (3,-.9)     [gp2 root] {i-1};
		\draw
		    (c) edge[->, gp2red] (a)
		;
		\node (A) at (2.5,-.42)   {\tiny{1}};
		\node (B) at (3.5,-.42)   {\tiny{2}};
		\node (C) at (3, -1.32)   {\tiny{3}};
		\end{tikzpicture}}
	
	& 
	
	\adjustbox{valign=t}{\begin{tikzpicture}
	    \node (a) at (0,0) [gp2 red node] {x};
		\node (b) at (1,0) [gp2 red node] {y};
		\node (c) at (.5,-.9)     [gp2 root] {i};
		\draw
		    (a) edge[->, gp2red] (b)
		    (c) edge[->, gp2red] (b)
		;
		\node (A) at (0,-.42)   {\tiny{1}};
		\node (B) at (1,-.42)   {\tiny{2}};
		\node (C) at (.5, -1.32)   {\tiny{3}};
		
		\node (c) at (1.75,-.5)     {$\Rightarrow$};
		
		\node (a) at (2.5,0) [gp2 red node] {x};
		\node (b) at (3.5,0) [gp2 red node] {y};
		\node (c) at (3,-.9)     [gp2 root] {i};
		\draw
		    (a) edge[->, gp2red] (b)
		    (c) edge[->, gp2red] (a)
		;
		\node (A) at (2.5,-.42)   {\tiny{1}};
		\node (B) at (3.5,-.42)   {\tiny{2}};
		\node (C) at (3, -1.32)   {\tiny{3}};
	\end{tikzpicture}}
	
\normalsize			
\end{tabular}
\end{allintypewriter}

\end{minipage}
}

\caption{The procedure \texttt{Rewind}}
\label{fig:mst-rewind}
\end{figure}


The procedure \ttt{MarkForDeletion} determines whether the current tree needs to be removed from the list of trees or not. This needs to be done when the current tree is being merged with another tree in the procedure \ttt{GrowForest}. However, in a set of trees that are merged into one tree, exactly one of them needs to be kept as an entry in the list. This is done by exploiting the fact that exactly one of the green edges used to merge that set of trees must have been selected by two different trees. If none of the edges fulfil that condition, the merging would introduce a cycle. If multiple edges fulfil it, the trees are merged into a forest, and not a single tree. Hence the trees that select a previously selected edge are kept as an entry in the list. The rule \ttt{clean} easily detects these edges since their label is a list of their edge weight followed by a \ttt{0}.



The procedure \ttt{Rewind} returns the pointer to the beginning of the list of trees. On the way, it removes list entries marked for deletion with a red loop, and updates the pointer's label which represents the number of trees in the list.

\section{Empirical Performance Results}
\label{sec:performance}

\begin{figure}[ht]
    \centering
    
    \begin{subfigure}{.5\textwidth}
    
    \centering
		\includegraphics{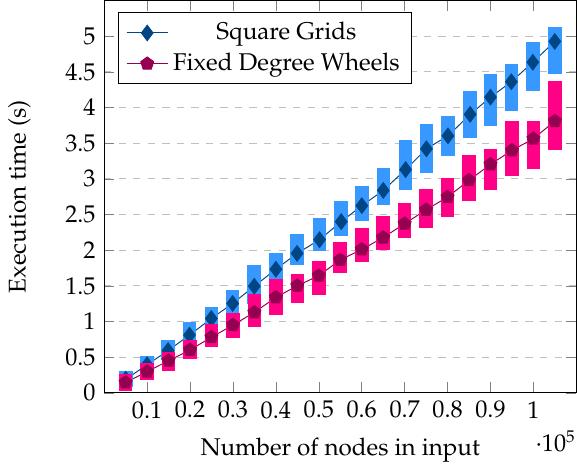}
        \caption{Bounded Degree Graphs}
        \label{fig:bnd-perf}
    \end{subfigure}%
    \begin{subfigure}{.5\textwidth}
    
    \centering
		\includegraphics{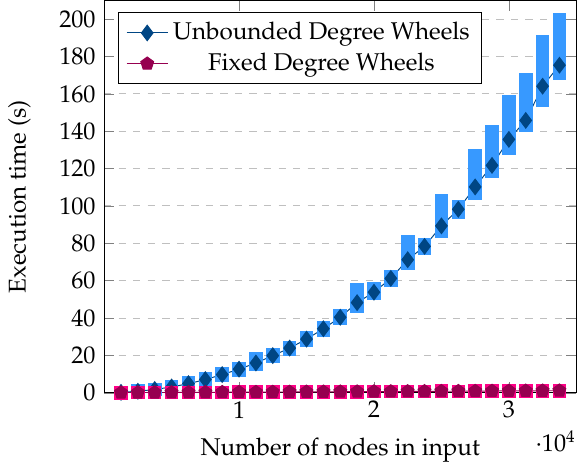}
        \caption{Wheels}
        \label{fig:wheels-perf}
    \end{subfigure}%

		\caption{Execution times of \texttt{mst-boruvka} with average}
		\label{fig:mst-perf}
	\end{figure}

On the graph classes we tested, time measurements as illustrated in Figure \ref{fig:mst-perf} show subquadratic growth on square grids and fixed degree wheels, and polynomial growth on unbounded degree wheels.

The execution time of the program \texttt{mst-boruvka} has been measured on square grids, fixed degree wheels, and undbounded degree wheels. The $k$\textsuperscript{th} \emph{square grid} is a $k \times k$ grid graph as depicted in Figure \ref{fig:ex-grid}. Figure \ref{fig:ex-wheel} depicts a wheel graph with $8$ spokes. The $k$\textsuperscript{th} \emph{fixed degree wheel} is a wheel graph with $16$ spokes, each of which consist of a path graph with $k$ edges. The $k$\textsuperscript{th} \emph{undbounded degree wheel} is a wheel graph with $k$ spokes.

The edge weights of the input graphs are randomly generated integers between $1$ and $1000$. The number of nodes of the square grids and fixed degree wheels ranges up to over $100000$, and that of the undbounded degree wheels to almost $35000$. For each graph of a given size, the execution time depicted with shapes is the average execution time of \texttt{mst-boruvka} on copies of that graph with at least $20$ random weight distributions. The bars around those data points show the range between the minimum and maximum measured execution time for that graph. The extent of that range can be attributed to differing random weight distributions used for each time measurement. With a fixed weight distribution, that range is much smaller.

Figure \ref{fig:bnd-perf} shows that \texttt{mst-boruvka} is subquadratic and close to linear on fixed degree wheels and square grids. We expect the time complexity to be $\text{O}(m \log n)$, where $m$ is the number of edges and $n$ the number of nodes, akin to those of standard implementations of Boruvka's MST algorithm \cite{Skiena-2008}. However, proving it will be left as future work. Note that, in order to reach this time complexity in GP\texorpdfstring{\,}{ }2, the use of root nodes is necessary.

In Figure \ref{fig:wheels-perf}, \texttt{mst-boruvka} is seen to be of an order worse than $m \log n$ on unbounded degree wheels. In fact, we conjecture it to be quadratic. GP\texorpdfstring{\,}{ }2 programs that are non-destructive in that they preserve the input graph seem to require at least quadratic time on unbounded degree input graphs. For example, consider \texttt{MinWithN!} seen in Figures \ref{fig:mst-findedge} and \ref{fig:mst-min}. In each case, it has to match a root (say $u$) and an adjacent non-root (say $v$) as long as possible. Assume that in the host graph, a root that is a valid match for $u$ has a linear number of adjacent nodes, all of which are a valid match for $v$. Assume that the first time \texttt{MinWithN} is called, the node with the highest edge weight is matched as $v$. The program only needs to check one node since every node is a valid match. Then assume that the second time \texttt{MinWithN} is called, the node with the next highest edge weight is matched. In the worst case, two nodes have to be checked for a valid match. Summing up the number of nodes that are checked if we continue this pattern, we get a sum of consecutive integers with a linear number of terms, which is quadratic. Hence the quadratic time complexity.

Furthermore, procedures that are based on depth-first search and preserve their input, such as the procedure \texttt{ColourBlue} in Figure \ref{fig:mst-blue}, have quadratic time complexity on unbounded degree graphs. The rule \texttt{blue\_back} looks for a dashed edge around the blue root. Assume the blue root has degree that is linear in the number of edges of a graph class the input graph belongs to. Only one of its adjacent edges can be dashed since the dashed edges form a path from the blue root to the origin of the DFS. Since there's only one valid match for the dashed edge, the rule application takes linear time. Every node of the input has to play the role of the blue root in \texttt{blue\_back} at some point.

The execution time on square grids is slower than that on fixed degree wheels by a constant factor. This is likely due to the fact that a large part of fixed degree wheels consists of path graphs, in which separate trees often share a minimum edge. So \ttt{MinWithN} is applied more often in fixed degree wheels than in square grids. Hence more rules (those of \ttt{MinWithN}) generally have to be called in square grids.

The time measurements were taken on a Lenovo Thinkpad T460 (2.4 GHz Intel Core i5, 16GB RAM) running Manjaro Linux, using the Python 3.8.3 time module. Exact figures of the time measurements can be found at \url{https://github.com/BrianCourtehoute/BrianCourtehoute.github.io/tree/master/PaperFiles/2020-06/Timings}, and the source program at \sloppy\url{https://github.com/BrianCourtehoute/BrianCourtehoute.github.io/blob/master/PaperFiles/2020-06/Code/mst-boruvka.gp2}.

\section{Conclusion and Future Work}

This paper features an implementation of Boruvka's algorithm for computing minimum spanning trees in the rule-based graph programming language GP\texorpdfstring{\,}{ }2. We have presented empirical evidence for its time efficiency on a few bounded degree graph classes that is in accordance with the known $\text{O}(m \log n)$ time complexity bound of Boruvka's algorithm implemented in conventional programming languages, where $m$ is the number of edges and $n$ the number of nodes. Furthermore, we have given empirical evidence for the program's non-linear time complexity on a graph class of unbounded degree.

The program is longer than its equivalent in conventional languages. The C implementation of Boruvka's algorithm presented by Sedgewick \cite{Sedgewick-1997,Sedgewick-2001} shown in Appendix \ref{app:boruvka-c} has $62$ lines of code (not accounting for lines only consisting of brackets). A textual representation of the GP\texorpdfstring{\,}{ }2 program has $330$ lines using a similar counting method. This is because, as an experimental language with a small syntax, GP\texorpdfstring{\,}{ }2 has no built-in data structures yet. Every data structure has to be implemented in the host graph itself. The procedures \ttt{Preprocess} and \ttt{Rewind} for instance only serve the purpose of creating and maintaining a list of trees.

Alternatively, one could omit counting the rule definitions in the GP\texorpdfstring{\,}{ }2 program, since rules are the basic operations of GP\texorpdfstring{\,}{ }2, and definitions of basic operations are not included in the line count of the C program. In that case, we count $30$ lines in the GP\texorpdfstring{\,}{ }2 program, which is much more comparable to the length of the C code.

Due to the large number of procedures and rules, it can be rather challenging to understand how the program \ttt{mst-boruvka} operates while reading it (there seems to be a trade-off between efficiency and readability). This goes against the GP\texorpdfstring{\,}{ }2 design philosophy of  facilitating formal reasoning, since proving soundness is not obvious. Indeed, giving a correctness proof in a formal proof system like in \cite{Wulandari-Plump-2020} would be a major undertaking. Hence the proof we plan to provide will not be in a formal proof system.

For the immediate future, we plan to write a longer version of this paper where we prove that $\text{O}(m \log n)$ is indeed a time bound of \ttt{mst-boruvka} on bounded degree graphs, and that the program indeed produces a minimum spanning tree of its input.

Another goal is to find more graph algorithms that can be implemented in GP\texorpdfstring{\,}{ }2 to reach their classical time bounds. We also plan to expand our technique for giving time measurements of a program's execution by comparing the timings of a GP\texorpdfstring{\,}{ }2 program to that of C code, and by using randomly generated inputs. Since the latter can produce wildly different timings however, it is not obvious how to do this in a sensible manner.

Finally, there is the open problem of how to create GP\texorpdfstring{\,}{ }2 programs that are efficient on unbounded degree graphs. There are GP\texorpdfstring{\,}{ }2 reduction programs in as yet unpublished work \cite{CampbellCourtehoutePlump20} that are efficient on arbitrary inputs. They operate by repeatedly removing nodes and edges from the host graph. This is a sensible approach for recognising whether a graph belongs to a certain graph class. However, when the purpose of a program is to produce a structure based on the input such as minimum spanning trees or topological sortings, this approach is not viable. It is not yet clear how to make these non-destructive programs efficient on arbitrary inputs.

\bibliographystyle{eptcs}
\bibliography{generic}

\appendix
\section{Boruvka's Algorithm in C}
\label{app:boruvka-c}

In this appendix, we include Sedgewick's C implementation of Boruvka's algorithm for the sake of comparison \cite{Sedgewick-1997,Sedgewick-2001}.

Listing \ref{lst:boruvka} contains the program computing a minimum spanning tree \cite{Sedgewick-2001}, which uses Union-Find as well as an adjacency list implementation of weighted graphs.

\begin{lstlisting}[caption=Boruvka's Algorithm, label=lst:boruvka]
Edge nn[maxV], a[maxE];
void GRAPHmstE(Graph G, Edge mst[])
  { int h, i, j, k, v, w, N; Edge e;
    int E = GRAPHedges(a, G);
    for (UFinit(G->V); E != 0; E = N)
      {
        for (k = 0; k < G->V; k++) 
          nn[k] = EDGE(G->V, G->V, maxWT);
        for (h = 0, N = 0; h < E; h++)
          {
            i = find(a[h].v); j = find(a[h].w); 
            if (i == j) continue;
            if (a[h].wt < nn[i].wt) nn[i] = a[h];
            if (a[h].wt < nn[j].wt) nn[j] = a[h];
            a[N++] = a[h]; 
          }
        for (k = 0; k < G->V; k++) 
          {
            e = nn[k]; v = e.v; w = e.w;
            if ((v != G->V) && !UFfind(v, w)) 
              { UFunion(v, w); mst[k] = e; }
          }
      }
  }
\end{lstlisting}

Listing \ref{lst:uf-int} shows the interface of Union-Find \cite{Sedgewick-1997}.

\begin{lstlisting}[caption=Union-Find ADT interface, label=lst:uf-int]
void UFinit(int);
int UFfind(int, int);
int UFunion(int, int);
\end{lstlisting}

Union-Find itself is shown in Listing \ref{lst:uf-imp} \cite{Sedgewick-1997}.

\begin{lstlisting}[caption=Union-Find ADT implementation, label=lst:uf-imp]
#include <stdlib.h>
#include "UF.h"
static int *id, *sz;
void UFinit(int N)
  { int i;
    id = malloc(N*sizeof(int)); 
    sz = malloc(N*sizeof(int)); 
    for (i = 0; i < N; i++) 
      { id[i] = i; sz[i] = 1; }
  }
int find(int x)
  { int i = x; 
    while (i != id[i]) i = id[i]; return i; }
int UFfind(int p, int q)
  { return (find(p) == find(q)); }
int UFunion(int p, int q)
  { int i = find(p), j = find(q);
    if (i == j) return;
    if (sz[i] < sz[j])
         { id[i] = j; sz[j] += sz[i]; }
    else { id[j] = i; sz[i] += sz[j]; }
  }
\end{lstlisting}

Listing \ref{lst:adjacency} contains the implementation of weighted graphs using adjacency lists \cite{Sedgewick-2001}.

\begin{lstlisting}[caption=Weighted-graph ADT (adjacency lists), label=lst:adjacency]
#include "GRAPH.h"
typedef struct node *link;
struct node { int v; double wt; link next; };
struct graph { int V; int E; link *adj; };
link NEW(int v, double wt, link next)
  { link x = malloc(sizeof *x);
    x->v = v; x->wt = wt; x->next = next;     
    return x;                         
  }
Graph GRAPHinit(int V)
  { int i;
    Graph G = malloc(sizeof *G);
    G->adj = malloc(V*sizeof(link));
    G->V = V; G->E = 0;
    for (i = 0; i < V; i++) G->adj[i] = NULL;
    return G;
  }
void GRAPHinsertE(Graph G, Edge e)
  { link t; 
    int v = e.v, w = e.w;
    if (v == w) return;
    G->adj[v] = NEW(w, e.wt, G->adj[v]);
    G->adj[w] = NEW(v, e.wt, G->adj[w]);
    G->E++;
  }
\end{lstlisting}

\end{document}

%% file: macros.tex
\newcommand{\ttt}{\texttt}

\newcommand{\mtt}{\mathtt}

\newcommand{\dder}{\Rightarrow}

%% file: main.bbl
\begin{thebibliography}{10}
\providecommand{\bibitemdeclare}[2]{}
\providecommand{\surnamestart}{}
\providecommand{\surnameend}{}
\providecommand{\urlprefix}{Available at }
\providecommand{\url}[1]{\texttt{#1}}
\providecommand{\href}[2]{\texttt{#2}}
\providecommand{\urlalt}[2]{\href{#1}{#2}}
\providecommand{\doi}[1]{doi:\urlalt{http://dx.doi.org/#1}{#1}}
\providecommand{\bibinfo}[2]{#2}

\bibitemdeclare{article}{Agrawal-Karsai-Neema-Shi-Vizhanyo06a}
\bibitem{Agrawal-Karsai-Neema-Shi-Vizhanyo06a}
\bibinfo{author}{Aditya \surnamestart Agrawal\surnameend},
  \bibinfo{author}{Gabor \surnamestart Karsai\surnameend},
  \bibinfo{author}{Sandeep \surnamestart Neema\surnameend},
  \bibinfo{author}{Feng \surnamestart Shi\surnameend} \&
  \bibinfo{author}{Attila \surnamestart Vizhanyo\surnameend}
  (\bibinfo{year}{2006}): \emph{\bibinfo{title}{The design of a language for
  model transformations}}.
\newblock {\sl \bibinfo{journal}{Software and System Modeling}}
  \bibinfo{volume}{5}(\bibinfo{number}{3}), pp. \bibinfo{pages}{261--288},
  \doi{10.1007/s10270-006-0027-7}.

\bibitemdeclare{inproceedings}{Arendt-Biermann-Jurack-Krause-Taentzer10a}
\bibitem{Arendt-Biermann-Jurack-Krause-Taentzer10a}
\bibinfo{author}{Thorsten \surnamestart Arendt\surnameend},
  \bibinfo{author}{Enrico \surnamestart Biermann\surnameend},
  \bibinfo{author}{Stefan \surnamestart Jurack\surnameend},
  \bibinfo{author}{Christian \surnamestart Krause\surnameend} \&
  \bibinfo{author}{Gabriele \surnamestart Taentzer\surnameend}
  (\bibinfo{year}{2010}): \emph{\bibinfo{title}{Henshin: Advanced Concepts and
  Tools for In-Place {EMF} Model Transformations}}.
\newblock In: {\sl \bibinfo{booktitle}{Model Driven Engineering Languages and
  Systems (MODELS 2010)}}, {\sl \bibinfo{series}{Lecture Notes in Computer
  Science}} \bibinfo{volume}{6394}, \bibinfo{publisher}{Springer}, pp.
  \bibinfo{pages}{121--135}, \doi{10.1007/978-3-642-16145-2\_9}.

\bibitemdeclare{phdthesis}{Bak15a}
\bibitem{Bak15a}
\bibinfo{author}{Christopher \surnamestart Bak\surnameend}
  (\bibinfo{year}{2015}): \emph{\bibinfo{title}{{GP\,2}: Efficient
  Implementation of a Graph Programming Language}}.
\newblock Ph.D. thesis, \bibinfo{school}{Department of Computer Science,
  University of York}.
\newblock \urlprefix\url{http://etheses.whiterose.ac.uk/12586/}.

\bibitemdeclare{inproceedings}{Bak-Plump12a}
\bibitem{Bak-Plump12a}
\bibinfo{author}{Christopher \surnamestart Bak\surnameend} \&
  \bibinfo{author}{Detlef \surnamestart Plump\surnameend}
  (\bibinfo{year}{2012}): \emph{\bibinfo{title}{Rooted Graph Programs}}.
\newblock In: {\sl \bibinfo{booktitle}{Proc.\ International Workshop on Graph
  Based Tools (GraBaTs 2012)}}, {\sl \bibinfo{series}{Electronic Communications
  of the EASST}}~\bibinfo{volume}{54}, \doi{10.14279/tuj.eceasst.54.780}.

\bibitemdeclare{inproceedings}{Bak-Plump16a}
\bibitem{Bak-Plump16a}
\bibinfo{author}{Christopher \surnamestart Bak\surnameend} \&
  \bibinfo{author}{Detlef \surnamestart Plump\surnameend}
  (\bibinfo{year}{2016}): \emph{\bibinfo{title}{Compiling Graph Programs to
  {C}}}.
\newblock In: {\sl \bibinfo{booktitle}{Proc.\ International Conference on Graph
  Transformation ({ICGT 2016})}}, {\sl \bibinfo{series}{LNCS}}
  \bibinfo{volume}{9761}, \bibinfo{publisher}{Springer}, pp.
  \bibinfo{pages}{102--117}, \doi{10.1007/978-3-319-40530-8\_7}.

\bibitemdeclare{article}{Cunyet-Khali-2001}
\bibitem{Cunyet-Khali-2001}
\bibinfo{author}{C{\"u}neyt~F. \surnamestart Bazlama\c{c}c{\i}\surnameend} \&
  \bibinfo{author}{Khalil~S. \surnamestart Hindi\surnameend}
  (\bibinfo{year}{2001}): \emph{\bibinfo{title}{Minimum-weight spanning tree
  algorithms A survey and empirical study}}.
\newblock {\sl \bibinfo{journal}{Computers \& Operations Research}}
  \bibinfo{volume}{28}(\bibinfo{number}{8}), pp. \bibinfo{pages}{767--785},
  \doi{10.1016/S0305-0548(00)00007-1}.

\bibitemdeclare{unpublished}{CampbellCourtehoutePlump20}
\bibitem{CampbellCourtehoutePlump20}
\bibinfo{author}{Graham \surnamestart Campbell\surnameend},
  \bibinfo{author}{Brian \surnamestart Courtehoute\surnameend} \&
  \bibinfo{author}{Detlef \surnamestart Plump\surnameend}:
  \emph{\bibinfo{title}{Fast Rule-Based Graph Programs}}.
\newblock \bibinfo{note}{Work in progress}.

\bibitemdeclare{inproceedings}{Campbell-Courtehoute-Plump-2019}
\bibitem{Campbell-Courtehoute-Plump-2019}
\bibinfo{author}{Graham \surnamestart Campbell\surnameend},
  \bibinfo{author}{Brian \surnamestart Courtehoute\surnameend} \&
  \bibinfo{author}{Detlef \surnamestart Plump\surnameend}
  (\bibinfo{year}{2019}): \emph{\bibinfo{title}{Linear-Time Graph Algorithms in
  {GP\,2}}}.
\newblock In: {\sl \bibinfo{booktitle}{Proceedings 8th Conference on Algebra
  and Coalgebra in Computer Science (CALCO 2019)}}, \bibinfo{series}{Leibniz
  International Proceedings in Informatics (LIPICS)},
  \bibinfo{publisher}{Schloss Dagstuhl-Leibniz-Zentrum fuer Informatik}, pp.
  \bibinfo{pages}{16:1--16:23}, \doi{10.4230/LIPIcs.CALCO.2019.16}.

\bibitemdeclare{article}{CampbellRomoPlump20}
\bibitem{CampbellRomoPlump20}
\bibinfo{author}{Graham \surnamestart Campbell\surnameend},
  \bibinfo{author}{Jack \surnamestart Rom\"o\surnameend} \&
  \bibinfo{author}{Detlef \surnamestart Plump\surnameend}
  (\bibinfo{year}{2020}): \emph{\bibinfo{title}{The Improved {GP\,2}
  Compiler}}.
\newblock {\sl \bibinfo{journal}{ArXiv e-prints}}
  \bibinfo{volume}{arXiv:2010.03993}.
\newblock \urlprefix\url{https://arxiv.org/abs/2010.03993}.
\newblock \bibinfo{note}{11 pages}.

\bibitemdeclare{inproceedings}{Courtehoute-Plump-2020}
\bibitem{Courtehoute-Plump-2020}
\bibinfo{author}{Brian \surnamestart Courtehoute\surnameend} \&
  \bibinfo{author}{Detlef \surnamestart Plump\surnameend}
  (\bibinfo{year}{2020}): \emph{\bibinfo{title}{A Fast Graph Program for
  Computing Minimum Spanning Trees}}.
\newblock In: {\sl \bibinfo{booktitle}{Proc. 11th International Workshop on
  Graph Computation Models (GCM 2020)}}, pp. \bibinfo{pages}{165--183}.
\newblock
  \urlprefix\url{http://www.cs.york.ac.uk/plasma/publications/pdf/Courtehoute-Plump.GCM.20.pdf}.
\newblock \bibinfo{note}{Pre-proceedings version of this paper}.

\bibitemdeclare{inproceedings}{Fernandez-Kirchner-Mackie-Pinaud14a}
\bibitem{Fernandez-Kirchner-Mackie-Pinaud14a}
\bibinfo{author}{Maribel \surnamestart Fern{\'a}ndez\surnameend},
  \bibinfo{author}{H{\'e}l{\`e}ne \surnamestart Kirchner\surnameend},
  \bibinfo{author}{Ian \surnamestart Mackie\surnameend} \&
  \bibinfo{author}{Bruno \surnamestart Pinaud\surnameend}
  (\bibinfo{year}{2014}): \emph{\bibinfo{title}{Visual Modelling of Complex
  Systems: Towards an Abstract Machine for {PORGY}}}.
\newblock In: {\sl \bibinfo{booktitle}{Proc.\ Computability in Europe (CiE
  2014)}}, {\sl \bibinfo{series}{Lecture Notes in Computer Science}}
  \bibinfo{volume}{8493}, \bibinfo{publisher}{Springer}, pp.
  \bibinfo{pages}{183--193}, \doi{10.1007/978-3-319-08019-2\_19}.

\bibitemdeclare{article}{Ghamarian-deMol-Rensink-Zambon-Zimakova12a}
\bibitem{Ghamarian-deMol-Rensink-Zambon-Zimakova12a}
\bibinfo{author}{Amir~Hossein \surnamestart Ghamarian\surnameend},
  \bibinfo{author}{Maarten \surnamestart de~Mol\surnameend},
  \bibinfo{author}{Arend \surnamestart Rensink\surnameend},
  \bibinfo{author}{Eduardo \surnamestart Zambon\surnameend} \&
  \bibinfo{author}{Maria \surnamestart Zimakova\surnameend}
  (\bibinfo{year}{2012}): \emph{\bibinfo{title}{Modelling and analysis using
  {GROOVE}}}.
\newblock {\sl \bibinfo{journal}{International Journal on Software Tools for
  Technology Transfer}} \bibinfo{volume}{14}(\bibinfo{number}{1}), pp.
  \bibinfo{pages}{15--40}, \doi{10.1007/s10009-011-0186-x}.

\bibitemdeclare{inproceedings}{Habel-Plump02c}
\bibitem{Habel-Plump02c}
\bibinfo{author}{Annegret \surnamestart Habel\surnameend} \&
  \bibinfo{author}{Detlef \surnamestart Plump\surnameend}
  (\bibinfo{year}{2002}): \emph{\bibinfo{title}{Relabelling in Graph
  Transformation}}.
\newblock In: {\sl \bibinfo{booktitle}{Proc.\ International Conference on Graph
  Transformation (ICGT 2002)}}, {\sl \bibinfo{series}{Lecture Notes in Computer
  Science}} \bibinfo{volume}{2505}, \bibinfo{publisher}{Springer}, pp.
  \bibinfo{pages}{135--147}, \doi{10.1007/3-540-45832-8\_12}.

\bibitemdeclare{article}{Jakumeit-Buchwald-Kroll10a}
\bibitem{Jakumeit-Buchwald-Kroll10a}
\bibinfo{author}{Edgar \surnamestart Jakumeit\surnameend},
  \bibinfo{author}{Sebastian \surnamestart Buchwald\surnameend} \&
  \bibinfo{author}{Moritz \surnamestart Kroll\surnameend}
  (\bibinfo{year}{2010}): \emph{\bibinfo{title}{{GrGen.NET} - The expressive,
  convenient and fast graph rewrite system}}.
\newblock {\sl \bibinfo{journal}{International Journal on Software Tools for
  Technology Transfer}} \bibinfo{volume}{12}(\bibinfo{number}{3--4}), pp.
  \bibinfo{pages}{263--271}, \doi{10.1007/s10009-010-0148-8}.

\bibitemdeclare{inproceedings}{Plump12a}
\bibitem{Plump12a}
\bibinfo{author}{Detlef \surnamestart Plump\surnameend} (\bibinfo{year}{2012}):
  \emph{\bibinfo{title}{The Design of {GP\,2}}}.
\newblock In: {\sl \bibinfo{booktitle}{Proc.\ Workshop on Reduction Strategies
  in Rewriting and Programming (WRS 2011)}}, {\sl \bibinfo{series}{Electronic
  Proceedings in Theoretical Computer Science}}~\bibinfo{volume}{82}, pp.
  \bibinfo{pages}{1--16}, \doi{10.4204/EPTCS.82.1}.

\bibitemdeclare{article}{Plump17a}
\bibitem{Plump17a}
\bibinfo{author}{Detlef \surnamestart Plump\surnameend} (\bibinfo{year}{2017}):
  \emph{\bibinfo{title}{From Imperative to Rule-based Graph Programs}}.
\newblock {\sl \bibinfo{journal}{Journal of Logical and Algebraic Methods in
  Programming}} \bibinfo{volume}{88}, pp. \bibinfo{pages}{154--173},
  \doi{10.1016/j.jlamp.2016.12.001}.

\bibitemdeclare{inproceedings}{Runge-Ermel-Taentzer11a}
\bibitem{Runge-Ermel-Taentzer11a}
\bibinfo{author}{Olga \surnamestart Runge\surnameend}, \bibinfo{author}{Claudia
  \surnamestart Ermel\surnameend} \& \bibinfo{author}{Gabriele \surnamestart
  Taentzer\surnameend} (\bibinfo{year}{2012}): \emph{\bibinfo{title}{{AGG 2.0}
  --- New Features for Specifying and Analyzing Algebraic Graph
  Transformations}}.
\newblock In: {\sl \bibinfo{booktitle}{Proc.\ Applications of Graph
  Transformations with Industrial Relevance (AGTIVE 2011)}}, {\sl
  \bibinfo{series}{Lecture Notes in Computer Science}} \bibinfo{volume}{7233},
  \bibinfo{publisher}{Springer}, pp. \bibinfo{pages}{81--88},
  \doi{10.1007/978-3-642-34176-2\_8}.

\bibitemdeclare{book}{Sedgewick-1997}
\bibitem{Sedgewick-1997}
\bibinfo{author}{Robert \surnamestart Sedgewick\surnameend}
  (\bibinfo{year}{1997}): \emph{\bibinfo{title}{Algorithms in C: Parts 1-4,
  Fundamentals, Data Structures, Sorting, and Searching}},
  \bibinfo{edition}{3rd} edition.
\newblock \bibinfo{publisher}{Addison-Wesley}.

\bibitemdeclare{book}{Sedgewick-2001}
\bibitem{Sedgewick-2001}
\bibinfo{author}{Robert \surnamestart Sedgewick\surnameend}
  (\bibinfo{year}{2001}): \emph{\bibinfo{title}{Algorithms in C, Part 5: Graph
  Algorithms}}, \bibinfo{edition}{3rd} edition.
\newblock \bibinfo{publisher}{Addison-Wesley}.

\bibitemdeclare{book}{Skiena-2008}
\bibitem{Skiena-2008}
\bibinfo{author}{Steven~S. \surnamestart Skiena\surnameend}
  (\bibinfo{year}{2008}): \emph{\bibinfo{title}{The Algorithm Design Manual}},
  \bibinfo{edition}{2nd} edition.
\newblock \bibinfo{publisher}{Springer}, \doi{10.1007/978-1-84800-070-4}.

\bibitemdeclare{inproceedings}{Wulandari-Plump-2020}
\bibitem{Wulandari-Plump-2020}
\bibinfo{author}{Gia \surnamestart Wulandari\surnameend} \&
  \bibinfo{author}{Detlef \surnamestart Plump\surnameend}
  (\bibinfo{year}{2020}): \emph{\bibinfo{title}{Verifying Graph Programs with
  First-Order Logic}}.
\newblock In: {\sl \bibinfo{booktitle}{Graph Computation Models (GCM 2020),
  Revised Selected Papers}}, {\sl \bibinfo{series}{Electronic Proceedings in
  Theoretical Computer Science}} \bibinfo{volume}{This volume}.

\end{thebibliography}
